\documentclass[12pt,preprint]{aastex}
\usepackage{graphicx}
\usepackage{amsfonts,amsmath,amssymb,mathrsfs}
\usepackage{color}

\newcommand{\be}{\begin{eqnarray}}
\newcommand{\ee}{\end{eqnarray}}

\begin{document}

\title{A code to compute the emission of thin accretion disks in non-Kerr space-times and test the nature of black hole candidates}

\author{Cosimo Bambi}
\affil{Arnold Sommerfeld Center for Theoretical Physics,\\
Ludwig-Maximilians-Universit\"at M\"unchen, D-80333 Munich, Germany}
\affil{Center for Field Theory and Particle Physics \& Department of Physics,\\ 
Fudan University, 200433 Shanghai, China}

\date{\today}

\begin{abstract}
Astrophysical black hole candidates are thought to be the Kerr black holes
predicted by General Relativity, but the actual nature of these objects has
still to be proven. The analysis of the electromagnetic radiation emitted by
a geometrically thin and optically thick accretion disk around a black hole 
candidate can provide information about the geometry of the space-time 
around the compact object and it can thus test the Kerr black hole hypothesis. 
In this paper, I present a code based on a ray-tracing approach and capable 
of computing some basic properties of thin accretion disks in space-times 
with deviations from the Kerr background. The code can be used to fit 
current and future X-ray data of stellar-mass black hole candidates and 
constrain possible deviations from the Kerr geometry in the spin 
parameter-deformation parameter plane.
\end{abstract}



\section{Introduction}

The predictions of General Relativity have been confirmed by experiments 
in Earth's gravitational field \citep{llr,gpb}, by spacecraft missions in 
the Solar System \citep{cassini}, and by accurate radio observations of 
binary pulsars \citep{wt05,pulsar} (for a general review, see e.g. \citet{will}). 
In all these environments, the gravitational field is weak, in the sense that 
one can write $g_{tt} = 1 + \phi$ with $|\phi| \ll 1$. The validity of the theory 
in the regime of strong gravity, when the approximation $|\phi| \ll 1$ breaks 
down, is instead still unexplored. The ideal laboratory to test strong gravitational 
fields is the space-time around astrophysical black hole (BH) candidates \citep{b1}.

In 4-dimensional General Relativity, uncharged BHs are described by the 
Kerr solution and are completely specified by two parameters: the mass, $M$, and 
the spin angular momentum, $J$ \citep{hair1,hair2}. A fundamental limit for a Kerr 
BH is the bound $|a| \le M$, where $a = J/M$ is the spin 
parameter\footnote{Throughout the paper, I use units in which 
$G_{\rm N} = c = 1$, unless stated otherwise.}. 
For $|a| > M$, there is no horizon, and the Kerr metric describes the 
gravitational field of a naked singularity, which is forbidden by the weak 
cosmic censorship conjecture \citep{wccc}. On the observational side, 
there are at least two classes of astrophysical BH candidates (for a review,
see e.g. \citet{n05}): stellar-mass compact objects in X-ray binary systems 
($M \approx 5 - 20$~$M_\odot$), and super-massive bodies at the center 
of every normal galaxy ($M \sim 10^5 - 10^9$~$M_\odot$). The measurement 
of the mass of the two classes of objects is robust, because obtained by 
dynamical methods and without any assumption about the nature of these 
objects. However, this is basically the only solid information we have. 
We think they are the Kerr BHs predicted by General Relativity 
because there is no alternative explanation in the framework of conventional 
physics: stellar-mass BH candidates are too heavy to be neutron star \citep{kb96},
while at least some of the super-massive objects in galactic nuclei are too heavy,
compact, and old to be clusters of non-luminous bodies \citep{maoz}.

The Kerr-nature of astrophysical BH candidates can potentially be tested with
already available X-ray data, by extending the two most popular techniques 
currently used by astronomers to estimate the spin parameter of these objects: 
the continuum-fitting method \citep{z97,li05,cfm} and the K$\alpha$-iron line
analysis \citep{iron1,iron2,iron3}. With the continuum-fitting method, one studies 
the thermal spectrum of geometrically thin and optically thick accretion disks: 
under the assumption of Kerr background and with independent measurements 
of the mass of the object, its distance from us, and the inclination angle of the 
disk, it is possible to infer the spin parameter $a$ and the mass accretion rate 
$\dot{M}$. The technique can be applied only to stellar-mass BH candidates, as 
the disk's temperature goes like $M^{-0.25}$ and the spectrum turns out to be 
in the keV-range for objects in X-ray binary systems, and in the UV for the 
super-massive bodies at the centers of galaxies. Relaxing the Kerr BH hypothesis, 
one can investigate the geometry of the space-time around the BH candidate 
\citep{bb11}. With the same spirit, one can analyze the broad K$\alpha$-iron
lines observed in both stellar-mass and super-massive BH candidates to test
their nature \citep{jpk}. Actually, one can usually get
only a constraint on a certain combination of the spin parameter and of the 
deviations from the Kerr background. In order to really test the nature of the 
compact object, at least two independent measurements are necessary 
\citep{b12a}.

With this paper, I am going to present a code designed to compute the thermal 
emission of a geometrically thin and optically thick accretion disk around a
compact object characterized by a mass, an arbitrary value of the spin parameter
(that is, no subject to the bound $|a| \le M$, valid only for Kerr BHs \citep{b11a1,
b11a2}), and by one (or more) ``deformation parameter(s)'', measuring possible 
deviations from the Kerr geometry. The code is based on a ray-tracing method
and takes all the relativistic effects into account, as well as some important 
astrophysical effects. It thus includes significant improvements with respect to
the code used in \citet{bb11} and can analyze real X-ray data providing the most 
reliable test we can get with current knowledge. 
The analysis of the soft X-ray component of specific sources and the corresponding
constraints on the spin parameter-deformation parameter plane will be presented 
in a forthcoming paper \citep{bgs}. The code can also be used to compute other
disk's properties, like its direct image and its light curve during an eclipse, which 
may be observed with future X-ray experiments and can provide additional 
details about the actual nature of a BH candidate.

\section{Thermal emission of thin accretion disks}

Geometrically thin and optically thick accretion disks are described by the
Novikov-Thorne model~\citep{nt1,nt2}, which is the relativistic generalization 
of the Shakura-Sunyaev model~\citep{ss-m}. 
Accretion is possible because viscous magnetic/turbulent stresses 
and radiation transport energy and angular momentum outwards. The 
model assumes that the disk is on the equatorial plane and that the disk's 
gas moves on nearly geodesic circular orbits. The time-averaged energy flux 
emitted from the surface of the disk is~\citep{nt2}
\be
\mathcal{F}(r) = \frac{\dot{M}}{4 \pi M^2} F(r) \, ,
\ee
where $F(r)$ is the dimensionless function
\be
F(r) = - \frac{\partial_r \Omega}{(E - \Omega L)^2} 
\frac{M^2}{\sqrt{-G}}
\int_{r_{\rm in}}^{r} (E - \Omega L) 
(\partial_\rho L) d\rho \, .
\nonumber\\
\ee
$E$, $L$, and $\Omega$ are, respectively, the conserved specific energy, 
the conserved axial-component of the specific angular momentum, and 
the angular velocity for equatorial circular geodesics (in a generic stationary
and axisymmetric space-time, these quantities can be computed as described,
for instance, in Appendix~B of \citet{bb11}); $G = - \alpha^2 g_{rr} g_{\phi\phi}$ 
is the determinant of the near equatorial plane metric, where $\alpha^2 =
g_{t\phi}^2/g_{\phi\phi} - g_{tt}$ is the lapse function; $r_{\rm in}$ is the 
inner radius of the accretion disk and is assumed to be the radius of the
innermost stable circular orbit (ISCO).

Since the disk is in thermal equilibrium, the emission is blackbody-like 
and we can define an effective temperature $T_{\rm eff} (r)$ from the 
relation $\mathcal{F}(r) = \sigma T^4_{\rm eff}$, where $\sigma$ is the 
Stefan-Boltzmann constant. Actually, the disk's temperature near the inner 
edge of the disk can be high, up to $\sim 10^7$~K for stellar-mass BH candidates, 
and non-thermal effects are non-negligible. That is usually taken into account
by introducing the color factor (or hardening factor) $f_{\rm col}$. The
color temperature is $T_{\rm col} (r) = f_{\rm col} T_{\rm eff}$ and the
local specific intensity of the radiation emitted by the disk is
\be\label{eq-i-bb}
I_{\rm e}(\nu_{\rm e}) = \frac{2 h \nu^3_{\rm e}}{c^2} \frac{1}{f_{\rm col}^4} 
\frac{\Upsilon}{\exp\left(\frac{h \nu_{\rm e}}{k_{\rm B} T_{\rm col}}\right) - 1} \, ,
\ee
where $\nu_{\rm e}$ is the photon frequency, 
$h$ is the Planck's constant, $c$ is the speed of light, $k_{\rm B}$ is 
the Boltzmann constant, and $\Upsilon$ is a function of the angle between 
the wavevector of the photon emitted by the disk and the normal of the
disk surface, say $\xi$. The two most common options are $\Upsilon = 1$
(isotropic emission) and $\Upsilon = \frac{1}{2} + \frac{3}{4} \cos\xi$ 
(limb-darkened emission).

\begin{table}
\centering
\begin{tabular}{lcp{9cm}}
& \hspace{0.5cm} & \\
\hline \hline
Geometry of the background & & Johannsen-Psaltis space-time with three 
parameters (mass $M$, spin parameter $a$, deformation parameter 
$\epsilon_3$). When $\epsilon_3 = 0$, the geometry of the space-time 
reduces exactly to the Kerr solution. No restrictions on the values of $a/M$ 
and $\epsilon_3$ \\
Relativistic effects & & All relativistic effects are included. Ray-tracing technique used \\
Self-irradiation & & Not included \\
Non-zero torque at $r_{\rm in}$ & & Not included \\
Indirect images & & Not included \\
Color factor $f_{\rm col}$ & & Constant. Set by the user \\
Radiation emission $\Upsilon$ & & Isotropic or limb-darkened \\
\hline \hline
\end{tabular}
\caption{Basic features of the code to compute the observed spectrum of the disk. Indirect
images are the images formed by null geodesics penetrating the equatorial plane inside 
$r_{\rm in}$.
\label{tab}}
\end{table}

\begin{table}
\centering
\begin{tabular}{ccccccccc}
\hline \hline
$a/M$ & \hspace{0.5cm} & $\epsilon_3$ & \hspace{0.5cm} & $r_{\rm H}/M$ & \hspace{0.5cm} & $r_{\rm ISCO}/M$ & \hspace{0.5cm} & ISCO \\
\hline \hline
0.0 && 0 && 2 && 6 && MRS \\
0.5 && 0 && 1.8660 && 4.2330 && MRS \\
0.9 && 0 && 1.4359 && 2.3209 && MRS \\
0.999 && 0 && 1.0447 && 1.1818 && MRS \\
\hline
0.8 && 2 && 1.6 $(\theta = 0, \, \pi)$ && 1.9048 && MVS \\
0.9 && 1 && 1.4359 $(\theta = 0, \, \pi)$ && 1.7499 && MVS \\
\hline
0.9 && -1 && 1.6176 $(\theta = \pi/2)$ && 3.2304 && MRS \\
1.1 && -0.5 && 1.2801 $(\theta = \pi/2)$ && 2.1139 && MRS \\
1.2 && -1 && 1.3614 $(\theta = \pi/2)$ && 2.5036 && MRS \\
\hline \hline
\end{tabular}
\caption{Properties of the backgrounds shown in 
Figs.~\ref{f-kerr}-\ref{f-ecl}: spin parameter $a$ (first column),
deformation parameter $\epsilon_3$ (second column), radius of the event horizon 
$r_{\rm H}$ (third column), radius of the ISCO $r_{\rm ISCO}$ (fourth column), and 
stability of the ISCO (MRS/MVS = Marginally Radially/Vertically Stable; fifth column). 
When $\epsilon_3 \neq 0$, the radius of the event horizon in general depends on
the angle $\theta$ and the topology of the event horizon may be non-trivial
(see \citet{leo1}). For the cases $(a/M , \epsilon_3) = (0.8,2)$ and $(0.9,1)$,
there is no event horizon on the equatorial plane $\theta = \pi/2$.
\label{tab2}}
\end{table}

\begin{figure}
\begin{center}  
\includegraphics[type=pdf,ext=.pdf,read=.pdf,width=8cm]{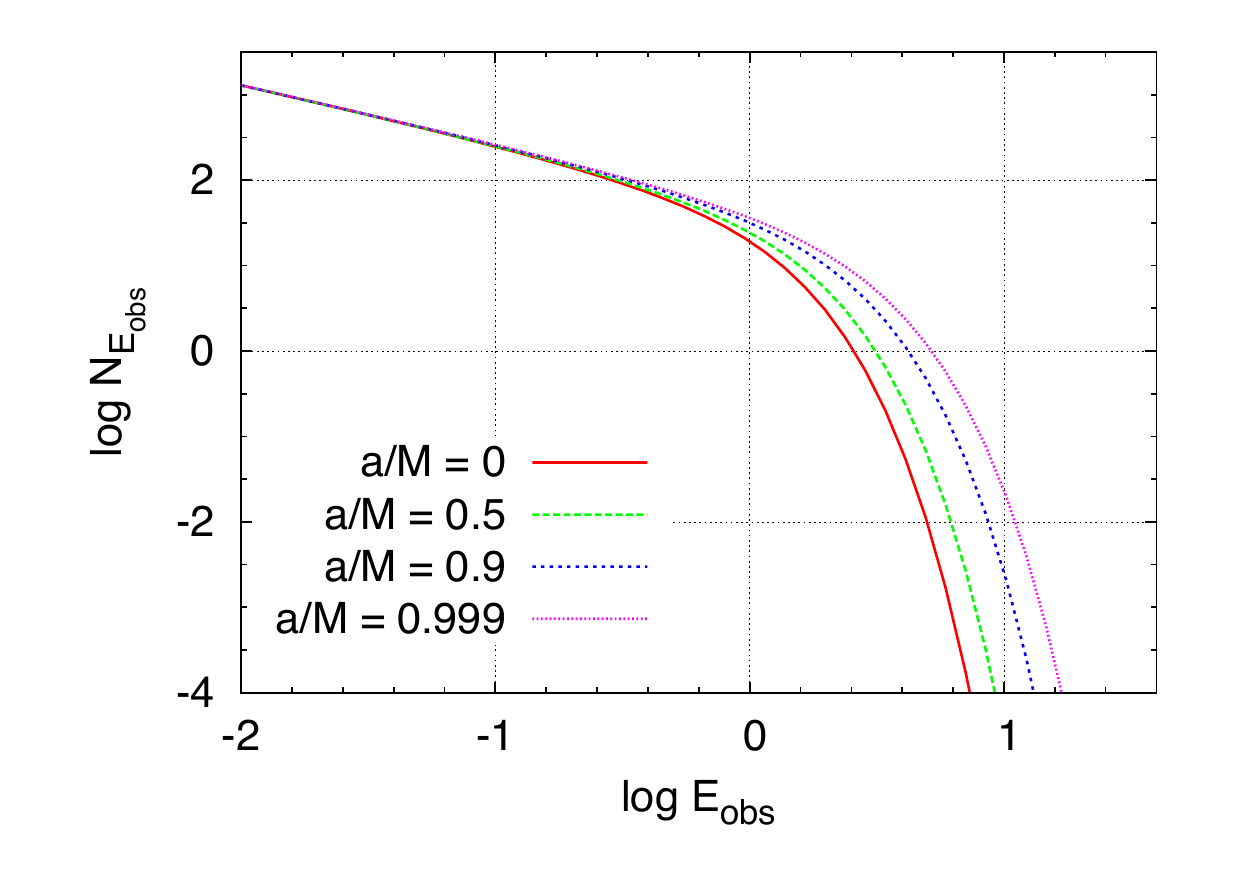}
\includegraphics[type=pdf,ext=.pdf,read=.pdf,width=8cm]{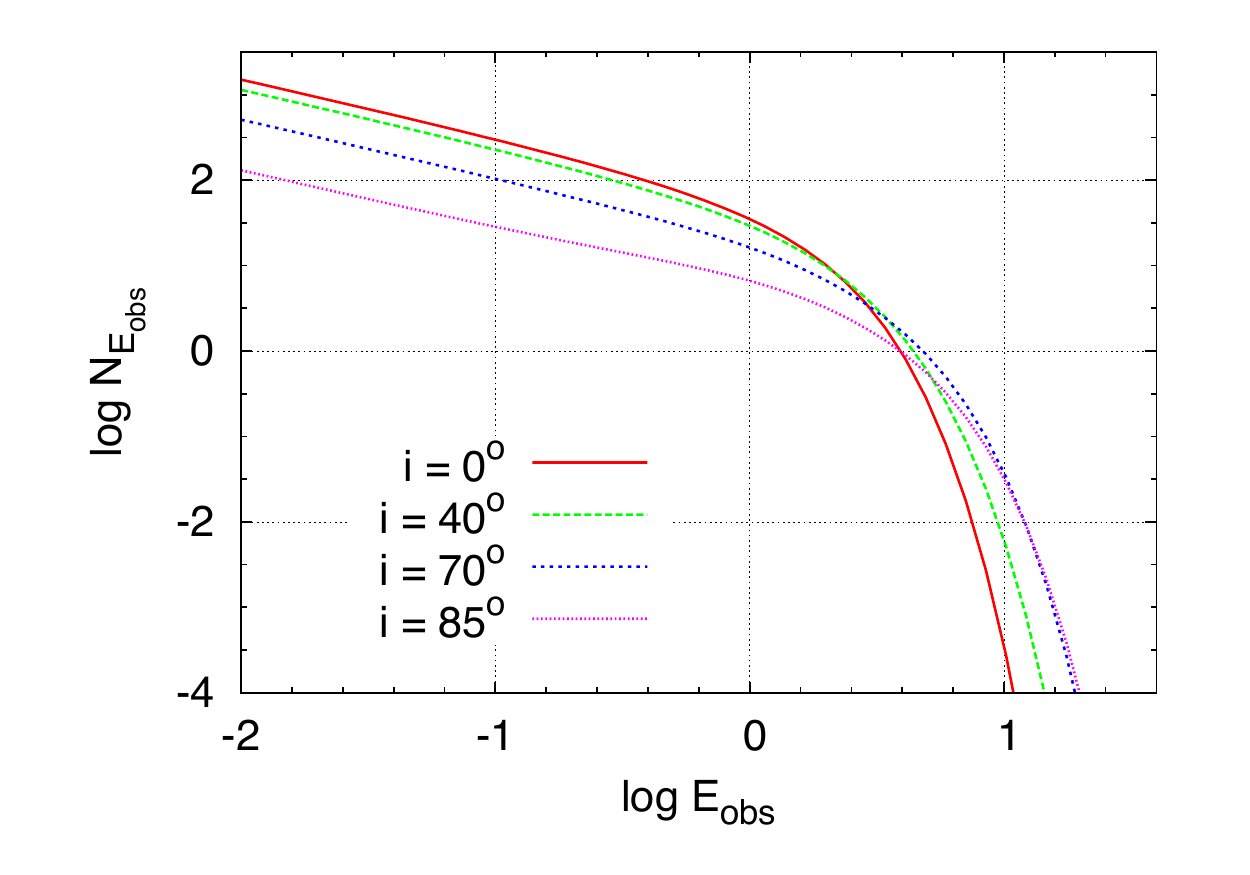}
\end{center}
\vspace{-0.6cm}
\caption{Observed thermal spectrum of a thin accretion disk in Kerr space-time. 
Left panel: $a/M = 0$, 0.5, 0.9, and 0.999 and viewing angle $i = 30^\circ$. 
Right panel: $a/M = 0.9$ and viewing angle $i = 0^\circ$, $40^\circ$, $70^\circ$, 
and $85^\circ$. The other parameters are $M = 10$~$M_\odot$, $\dot{M} = 
10^{19}$~g~s$^{-1}$, $D = 10$~kpc, $f_{\rm col} = 1$, and $\Upsilon = 1$. These two 
plots should be compared with the two panels of Fig.~5 in \citet{li05}. 
Flux density $N_{E_{\rm obs}}$ in $\gamma$ keV$^{-1}$ cm$^{-2}$ s$^{-1}$; 
photon energy $E_{\rm obs}$ in keV.}
\label{f-kerr}
\end{figure}

\begin{figure}
\begin{center}  
\includegraphics[type=pdf,ext=.pdf,read=.pdf,width=8cm]{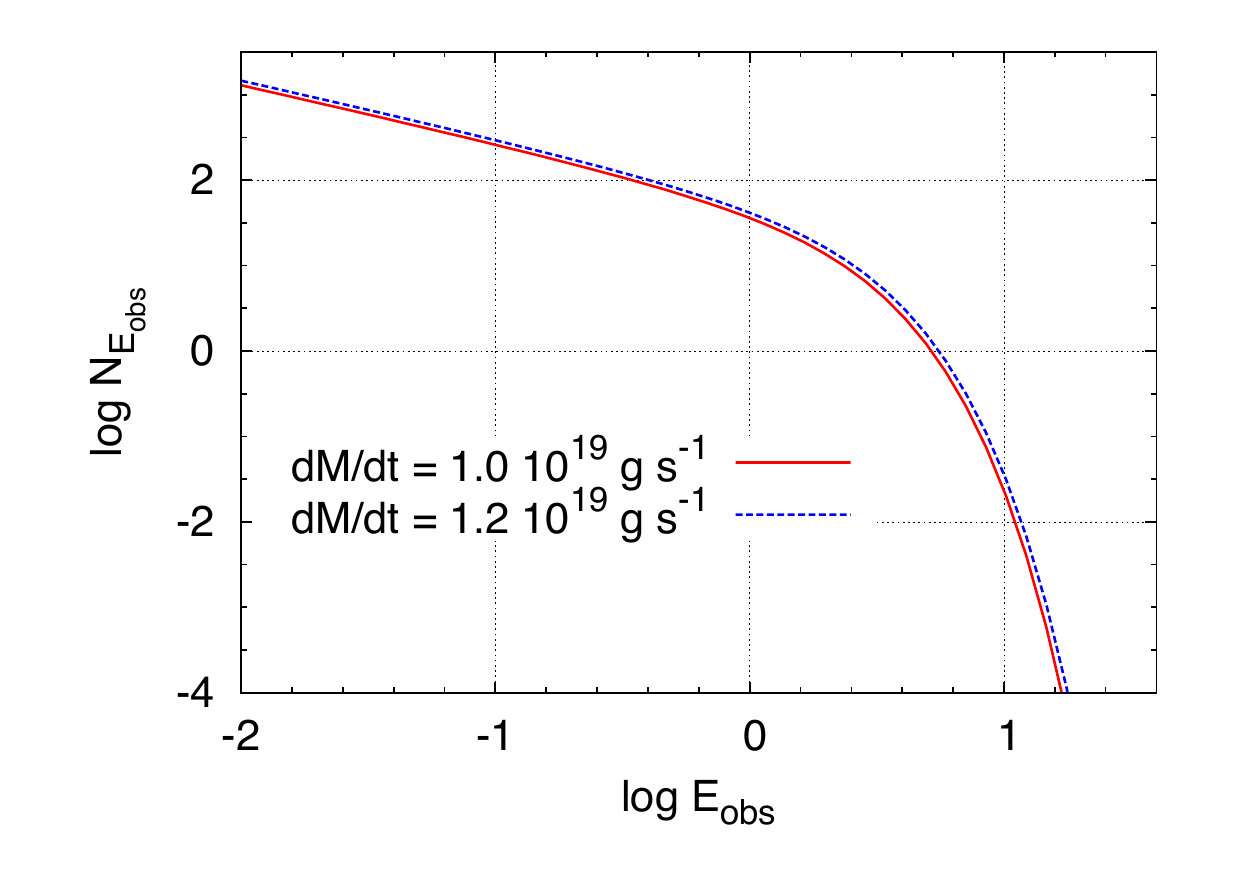}
\includegraphics[type=pdf,ext=.pdf,read=.pdf,width=8cm]{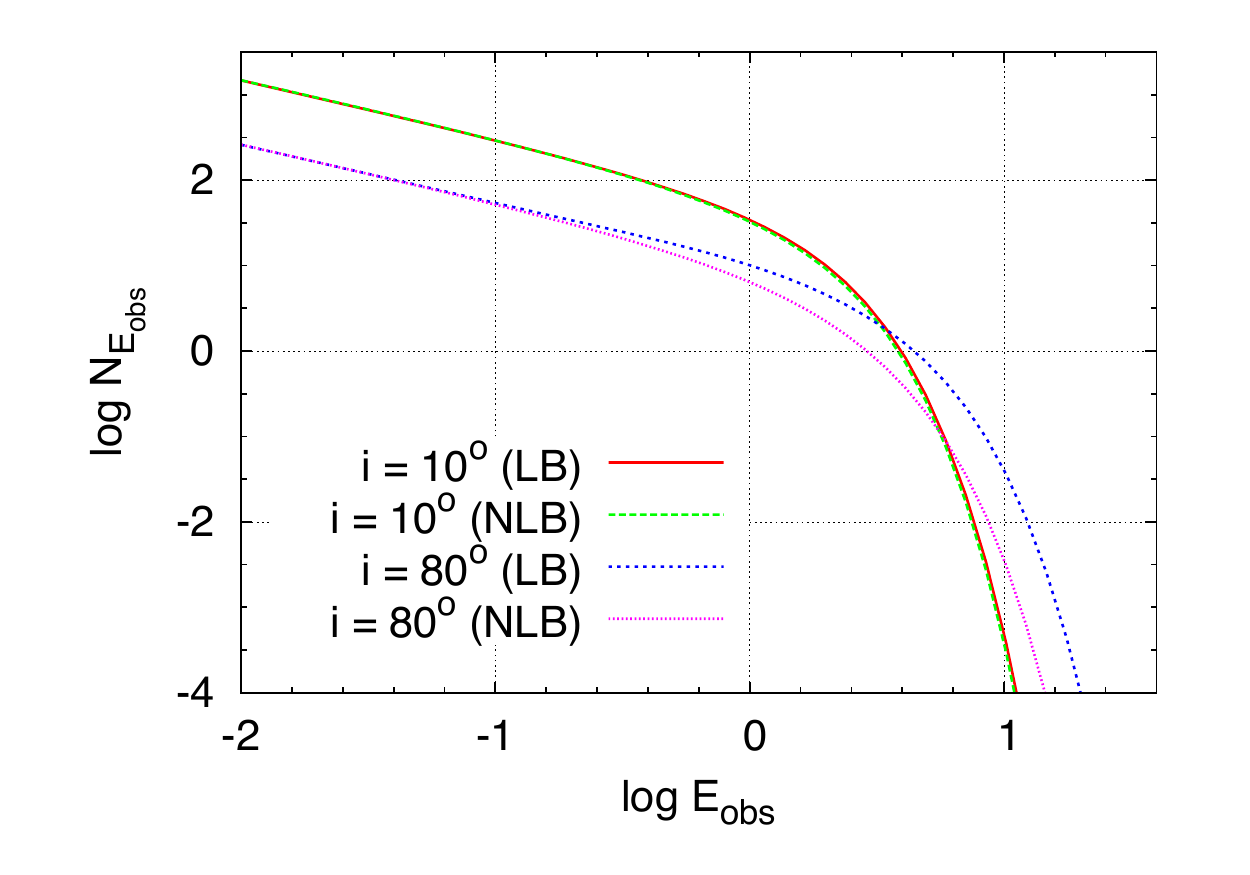}
\end{center}
\vspace{-0.6cm}
\caption{Left panel: Kerr space-time with $\dot{M} = 1.0 \cdot 10^{19}$~g~s$^{-1}$ 
and $1.2 \cdot 10^{19}$~g~s$^{-1}$. The other parameters are $M = 10$~$M_\odot$, 
$a/M = 0.999$, $i = 30^\circ$, $D = 10$~kpc, $f_{\rm col} = 1$, and $\Upsilon = 1$. 
The outer radius of the accretion disk is at $r_{\rm out} = 10^5$~$M$. Right 
panel: Effect of light bending on the observed spectrum in Kerr space-time for a viewing 
angle $i = 10^\circ$ and $80^\circ$. LB/NLB = light bending/no light bending.  The 
other parameters are $M = 10$~$M_\odot$, $a/M = 0.9$, $\dot{M} = 1.0 \cdot 
10^{19}$~g~s$^{-1}$, $D = 10$~kpc, $f_{\rm col} = 1$, and $\Upsilon = 1$. 
The outer radius of the accretion disk is at $r_{\rm out} = 10^5$~$M$. Flux density 
$N_{E_{\rm obs}}$ in $\gamma$ keV$^{-1}$ cm$^{-2}$ s$^{-1}$; photon energy 
$E_{\rm obs}$ in keV.}
\label{f-mdot}
\end{figure}

\subsection{Calculation method}

The calculation of the thermal spectrum of a thin accretion disk has
been extensively discussed in the literature; see e.g.~\citet{li05} 
and references therein. The spectrum can be conveniently written in terms of 
the photon flux number density as measured by a distant observer, 
$N_{E_{\rm obs}}$. A quite common approximation
in the literature is to neglect the effect of light bending, which is good for 
small disk's inclination angles (face-on disks). In this case, 
$N_{E_{\rm obs}}$ is:
\be\label{eq-n1}
N_{E_{\rm obs}} &=& \frac{1}{E_{\rm obs}} \int I_{\rm obs}(\nu) d \Omega_{\rm obs} = 
\nonumber\\ &=&
\frac{1}{E_{\rm obs}} \int g^3 I_{\rm e}(\nu_{\rm e}) d \Omega_{\rm obs} = 
\nonumber\\ &=& 
A_1 \left(\frac{E_{\rm obs}}{\rm keV}\right)^2 \cos i
\int_{\rm disk} \frac{1}{M^2}\frac{\Upsilon \sqrt{-G} dr d\phi}{\exp\left[\frac{A_2}{g F^{1/4}} 
\left(\frac{E_{\rm obs}}{\rm keV}\right)\right] - 1} \, ,
\ee
where $I_{\rm obs}$, $E_{\rm obs}$, and $\nu$ are, respectively, the specific
intensity of the radiation, the photon energy, and the photon frequency measured
by the distant observer. $d\Omega_{\rm obs} \approx \cos i \sqrt{-G} dr d\phi / D^2$ 
is the element of the solid angle subtended by the image of the disk on the 
observer's sky, $i$ is the viewing angle of the observer, and $D$ is the distance of 
the source. $g$ is the redshift factor
\be\label{eq-red}
g = \frac{E_{\rm obs}}{E_{\rm e}} = \frac{\nu}{\nu_{\rm e}} = 
\frac{k_\alpha u^{\alpha}_{\rm obs}}{k_\beta u^{\beta}_{\rm e}}\, ,
\ee
where $E_{\rm e} = h \nu_{\rm e}$,
$k^\alpha$ is the 4-momentum of the photon, $u^{\alpha}_{\rm obs} = (-1,0,0,0)$ 
is the 4-velocity of the distant observer, and $u^{\alpha}_{\rm e} = (u^t_{\rm e},0,0,
\Omega u^t_{\rm e})$ is the 4-velocity of the emitter.  
$I_{\rm e}(\nu_{\rm e})/\nu_{\rm e}^3 = I_{\rm obs} (\nu_{\rm obs})/\nu^3$ follows 
from the Liouville's theorem. $A_1$ and $A_2$ are 
given by (for the sake of clarity, here I show explicitly $G_{\rm N}$ and $c$)
\be
A_1 &=&  
\frac{2 \left({\rm keV}\right)^2}{f_{\rm col}^4} 
\left(\frac{G_{\rm N} M}{c^3 h D}\right)^2 = 
\nonumber\\ &=& 
\frac{0.07205}{f_{\rm col}^4} 
\left(\frac{M}{M_\odot}\right)^2 
\left(\frac{\rm kpc}{D}\right)^2 \, 
{\rm \gamma \, keV^{-1} \, cm^{-2} \, s^{-1}} \, , \nonumber\\
A_2 &=&  
\left(\frac{\rm keV}{k_{\rm B} f_{\rm col}}\right) 
\left(\frac{G_{\rm N} M}{c^3}\right)^{1/2}
\left(\frac{4 \pi \sigma}{\dot{M}}\right)^{1/4} = 
\nonumber\\ &=& 
\frac{0.1331}{f_{\rm col}} 
\left(\frac{\rm 10^{18} \, g \, s^{-1}}{\dot{M}}\right)^{1/4}
\left(\frac{M}{M_\odot}\right)^{1/2} \, .
\ee
Using the normalization condition
$g_{\mu\nu}u^{\mu}_{\rm e}u^{\nu}_{\rm e} = -1$, one finds
\be
u^t_{\rm e} = - \frac{1}{\sqrt{-g_{tt} - 2 g_{t\phi} \Omega - g_{\phi\phi} \Omega^2}} \, ,
\ee
and therefore
\be\label{eq-red-g}
g = \frac{\sqrt{-g_{tt} - 2 g_{t\phi} \Omega - g_{\phi\phi} \Omega^2}}{1 + 
\lambda \Omega} \, ,
\ee
where $\lambda = k_\phi/k_t$ is a constant of the motion along the photon path.
Neglecting the effect of light bending, $\lambda = r \sin\phi \sin i$, where $r$ and $\phi$
are the coordinates on the equatorial plane of the emitter.
Doppler boosting, gravitational redshift, and frame dragging are entirely encoded 
in the redshift factor $g$.

The effect of light bending can be included by using a ray-tracing approach. 
The initial conditions $(t_0, r_0, \theta_0, \phi_0)$ for the photon with Cartesian 
coordinates $(X,Y)$ on the image plane of the distant observer are given 
by \citep{jp10}
\be
t_0 &=& 0 \, , \\
r_0 &=& \sqrt{X^2 + Y^2 + D^2} \, , \\
\theta_0 &=& \arccos \frac{Y \sin i + D \cos i}{\sqrt{X^2 + Y^2 + D^2}} \, , \\
\phi_0 &=& \arctan \frac{X}{D \sin i - Y \cos i} \, .
\ee
As the initial 3-momentum $\bf{k}_0$ must be perpendicular to the plane of the 
image of the observer, the initial conditions for the 4-momentum of the photon 
are \citep{jp10}
\be
k^r_0 &=& - \frac{D}{\sqrt{X^2 + Y^2 + D^2}} |\bf{k}_0| \, , \\
k^\theta_0 &=& \frac{\cos i - D \frac{Y \sin i + D 
\cos i}{X^2 + Y^2 + D^2}}{\sqrt{X^2 + (D \sin i - Y \cos i)^2}} |\bf{k}_0| \, , \\
k^\phi_0 &=& \frac{X \sin i}{X^2 + (D \sin i - Y \cos i)^2} |\bf{k}_0| \, , \\
k^t_0 &=& \sqrt{\left(k^r_0\right)^2 + r^2_0  \left(k^\theta_0\right)^2
+ r_0^2 \sin^2\theta_0  (k^\phi_0)^2} \, . 
\ee
In the numerical calculation, the observer is located at $D = 10^6$~$M$, which 
is far enough to assume that the background geometry is flat and therefore
$k^t_0$ can be inferred from the condition $g_{\mu\nu}k^\mu k^\nu = 0$ with
the metric tensor of a flat space-time. The photon trajectory is numerically 
integrated backwards in time to the point of the photon emission on the 
accretion disk: in this way, we get the radial coordinate $r_{\rm e}$ at which 
the photon was emitted and the angle $\xi$ between the wavevector of the 
photon and the normal of the disk surface (necessary to compute $\Upsilon$).
Now $g = g(X,Y)$, as in Eq.~(\ref{eq-red-g}) everything depends on $r_{\rm e}$, 
except $\lambda$ that can be evaluated from the photon initial
conditions ($\lambda = k_\phi/k_t = r_0 \sin\theta_0 k^\phi_0 / k^t_0$).
The observerÕ's sky is divided into a number of small elements and the
ray-tracing procedure provides the observed flux density from each element; 
summing up all the elements, we get the total observed flux density of the 
disk. In the case of Kerr background, one can actually exploit the special 
properties of the Kerr solution and solve a simplified set of differential 
equations. That is not possible in a generic non-Kerr background and so 
the code solves the second-order photon geodesic equations of the space-time, 
by using the fourth-order Runge-Kutta-Nystr\"om method \citep{lund}. 
The photon flux number density is given by 
\be\label{eq-n2}
N_{E_{\rm obs}} &=&
\frac{1}{E_{\rm obs}} \int I_{\rm obs}(\nu) d \Omega_{\rm obs} = 
\frac{1}{E_{\rm obs}} \int g^3 I_{\rm e}(\nu_{\rm e}) d \Omega_{\rm obs} = 
\nonumber\\ &=& 
A_1 \left(\frac{E_{\rm obs}}{\rm keV}\right)^2
\int \frac{1}{M^2} \frac{\Upsilon dXdY}{\exp\left[\frac{A_2}{g F^{1/4}} 
\left(\frac{E_{\rm obs}}{\rm keV}\right)\right] - 1} \, ,
\ee
where $X$ and $Y$ are the coordinates of the position of the photon
on the sky, as seen by the distant observer; that is, $d\Omega_{\rm obs} = dX dY / D^2$. 
The basic features of the code are outlined in Tab.~\ref{tab}.

The results of the code in the Kerr background are summarized in 
Figs.~\ref{f-kerr} and \ref{f-mdot}, while Tab.~\ref{tab2} shows some fundamental
properties of these space-times. The two panels of Fig.~\ref{f-kerr}
should be compared with the ones in Fig.~5 of \citet{li05}. 
The agreement between the two sets of spectra is very good, 
except for the case $a/M = 0.999$ and inclination
angle $i = 30^\circ$, for which the difference is however small. 
The discrepancy is due to the fact that the present code does not include the 
effect of self-irradiation of the disk. As discussed in~\cite{li05},
the effect of self-irradiation of the disk, as well as a possible non-zero 
torque at the inner edge of the disk (also not included here, see Tab.~\ref{tab}),
can be ignored in the analysis of observational data, because
they can be absorbed by adjusting the mass accretion rate and the
spectral color factor of a zero torque model without returning
radiation. This is indeed our case, as shown in the left panel of
Fig.~\ref{f-mdot}, where we can see the spectra of a Kerr BH with
$a/M = 0.999$ and $i = 30^\circ$, respectively for $\dot{M} = 1.0 \cdot 10^{19}$
and $1.2 \cdot 10^{19}$~g~s$^{-1}$. The spectrum with higher
mass accretion rate and without self-irradiation of the disk is in agreement 
with the one calculated in \citet{li05} for $\dot{M} = 1.0 \cdot 
10^{19}$~g~s$^{-1}$ and with self-irradiation\footnote{With the standard
approach in a Kerr background, one has two parameters, $a$ and $\dot{M}$,
to be determined by fitting the soft X-ray component with the theoretical prediction
of the thermal spectrum of a thin disk. \citet{li05} shows that disk's self-irradiation
and small non-zero torque at the ISCO can be neglected without affecting the
measurement of $a$: one just gets a slightly different $\dot{M}$, which has not
physical implications. As the physical mechanisms responsible to these effects
are not peculiar of the Kerr geometry, the same conclusion should be true for
metrics with deviations from Kerr.}. The effect of light 
bending can be seen in the right panel in Fig.~\ref{f-mdot}, which
shows the observed spectra as computed respectively from 
Eq.~(\ref{eq-n1}) and Eq.~(\ref{eq-n2}). 
The difference in the observed spectra is significant for large
inclination angles (edge-on disks) and negligible in the opposite
case, when the angle between the observer and the symmetry 
axes of the system is small (face-on disks). In all these spectra, the inner edge
of the disk is at the ISCO radius (Novikov-Thorne model), while the outer
edge is assumed at $r_{\rm out} = 10^5$~$M$. The latter is large enough that 
for the energy range shown in Figs.~\ref{f-kerr}-\ref{f-limb} is equivalent to an
infinite value. The effect of $r_{\rm out}$ is indeed to introduce a different slope
of the spectrum at low energies, see e.g. Fig.~4 in \citet{bb11}.

\subsection{Non-Kerr space-times}

The Kerr-nature of astrophysical BH candidates can be tested by
studying the thermal spectrum of the accretion disk without the
assumption of the Kerr background. The basic idea is to consider 
a more general space-time, in which the compact object is
characterized by a mass $M$, a spin parameter $a$, and
a deformation parameter which measures possible deviation from
the Kerr geometry. When the deformation parameter is set to
zero, we have to recover exactly the Kerr solution. We can
then compare the theoretical predictions with observational data:
if the latter demand a zero deformation parameter, the Kerr 
BH hypothesis is verified; otherwise, we may conclude that
the object is not a Kerr BH.

Let us note that this approach can be used just to check if
the geometry of the space-time is described by the Kerr solution,
but it cannot really investigate the actual nature of the compact
object or the exact deviations from the Kerr background.
The analysis of the disk's spectrum can only probe the geometry of 
the space-time till the ISCO radius; we have no information
about the geometry at smaller radii and about the surface/horizon
of the BH candidate. Current (and near-future) data are not so good to map 
the space-time and a single deformation parameter is used to
figure out if the gravitational force is stronger 
or weaker than the one around a Kerr BH with the same mass and spin. 
Actually, the typical situation is even worse and 
we can infer only one parameter: if we assume the Kerr
background, we find the spin $a$, if we have also a deformation parameter,
we constrain some combination of $a$ and of the deformation parameter.
However, this degeneracy can be solved with an additional measurement 
\citep{b12a,b12b,b12c}.

When the computation of the properties of the thermal emission depends on
the geometry of the background, the code calls the function \verb|metric|.
It is thus easy to change the space-time. For the time being, the code uses
the Johannsen-Psaltis (JP) metric, which was explicitly proposed in \citet{jp11} 
to test the geometry around BH candidates. In Boyer-Lindquist coordinates, the 
metric is given by the line element \citep{jp11}
\be\label{eq-jp}
ds^2 &=& - \left(1 - \frac{2 M r}{\Sigma}\right) (1 + h) \, dt^2
- \frac{4 a M r \sin^2\theta}{\Sigma} (1 + h) \, dt \, d\phi
+ \frac{\Sigma (1 + h)}{\Delta + a^2 h \sin^2\theta } \, dr^2 + \nonumber\\
&& + \Sigma \, d\theta^2
+ \left[\sin^2\theta \left(r^2 + a^2 + \frac{2 a^2 M r \sin^2\theta}{\Sigma} \right) 
+ \frac{a^2 (\Sigma + 2 M r) \sin^4\theta}{\Sigma} h \right] d\phi^2 \, , 
\ee
where
\be
\Sigma &=& r^2 + a^2 \cos^2\theta \, , \nonumber\\
\Delta &=& r^2 - 2 M r + a^2 \, , \nonumber\\
h &=& \sum_{k = 0}^{\infty} \left(\epsilon_{2k} 
+ \frac{M r}{\Sigma} \epsilon_{2k+1} \right)
\left(\frac{M^2}{\Sigma}\right)^k \, .
\ee
This metric has an infinite number of deformation
parameters $\epsilon_i$ and the Kerr solution is recovered when
all the deformation parameters vanish. However,
in order to reproduce the correct Newtonian limit, we have to
impose $\epsilon_0 = \epsilon_1 = 0$, while $\epsilon_2$
is strongly constrained by Solar System experiments. As done
in other papers, it is usually sufficient to restrict the attention to the deformation
parameter $\epsilon_3$ and set to zero all the others.

\begin{figure}
\begin{center}  
\includegraphics[type=pdf,ext=.pdf,read=.pdf,width=8cm]{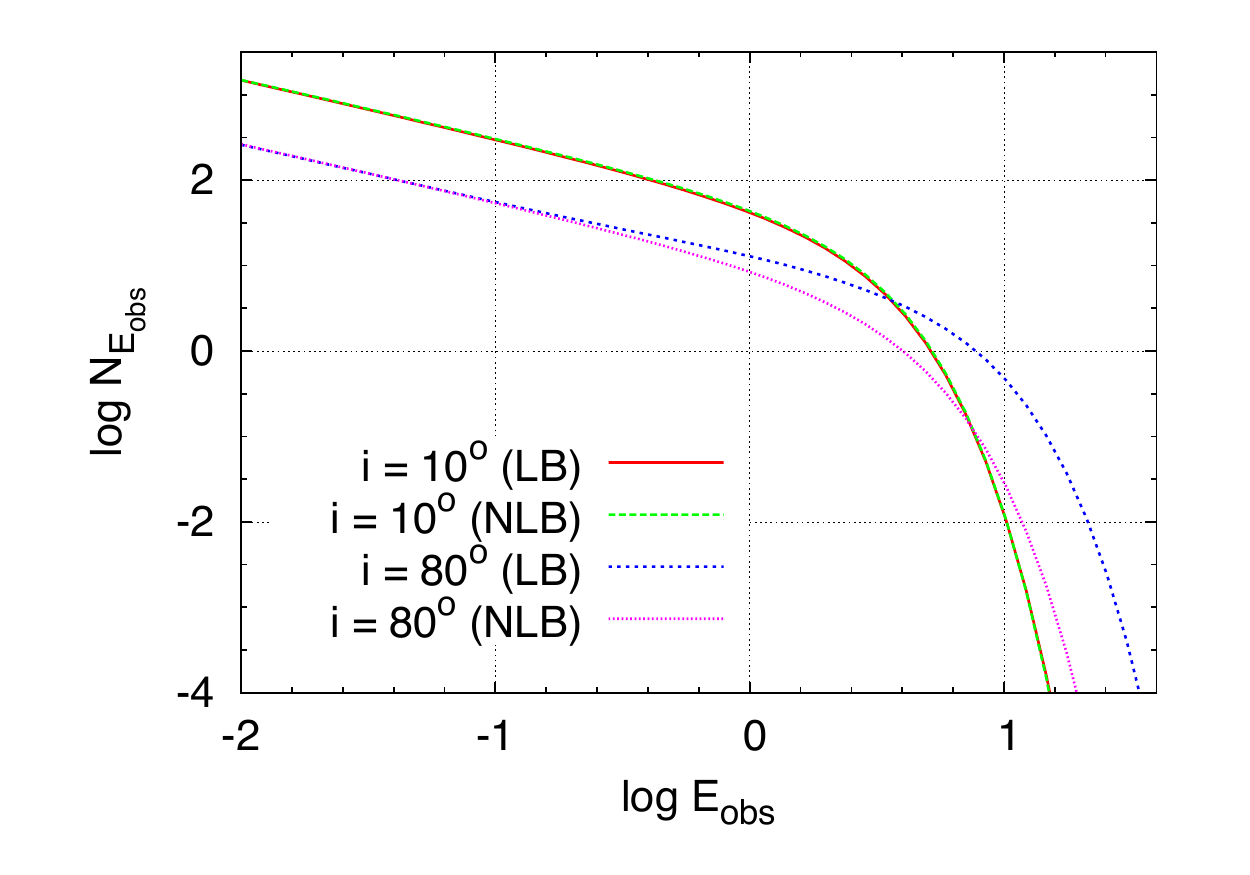}
\includegraphics[type=pdf,ext=.pdf,read=.pdf,width=8cm]{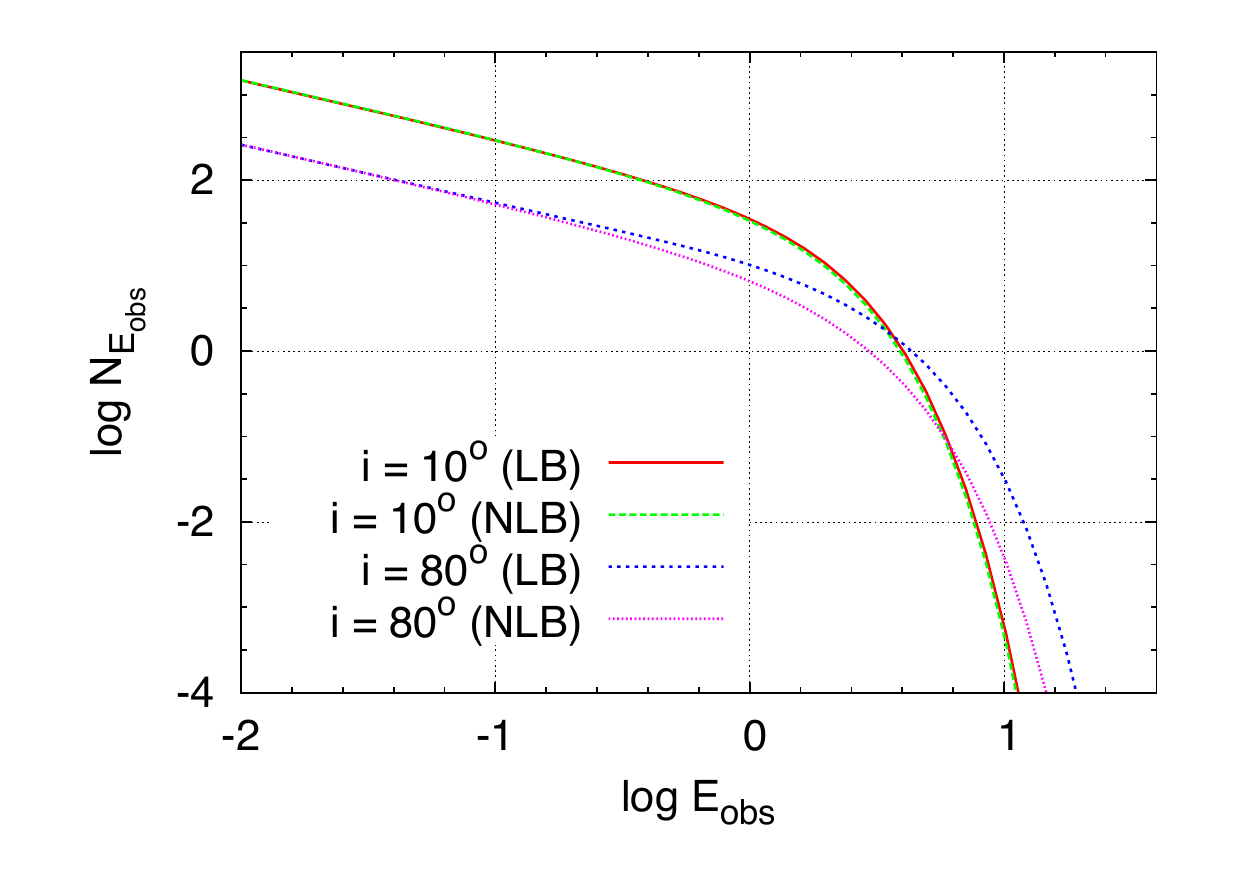}
\end{center}
\vspace{-0.6cm}
\caption{Effect of light bending on the observed spectrum in Johannsen-Psaltis 
space-time. Left panel: $a/M = 0.8$, $\epsilon_3 = 2$, and viewing angle 
$i = 10^\circ$ and $80^\circ$. Right panel: $a/M = 1.2$, $\epsilon_3 = - 1$, and 
viewing angle $i = 10^\circ$ and $80^\circ$. LB/NLB = light bending/no light bending. 
The other parameters are $M = 10$~$M_\odot$, $\dot{M} = 10^{19}$~g~s$^{-1}$, 
$D = 10$~kpc, $f_{\rm col} = 1$, and $\Upsilon = 1$. The outer radius of the 
accretion disk is at $r_{\rm out} = 10^5$~$M$. Flux density $N_{E_{\rm obs}}$ in 
$\gamma$ keV$^{-1}$ cm$^{-2}$ s$^{-1}$; photon energy $E_{\rm obs}$ in keV.}
\label{f-iobs}
\end{figure}

\begin{figure}
\begin{center}  
\includegraphics[type=pdf,ext=.pdf,read=.pdf,width=8cm]{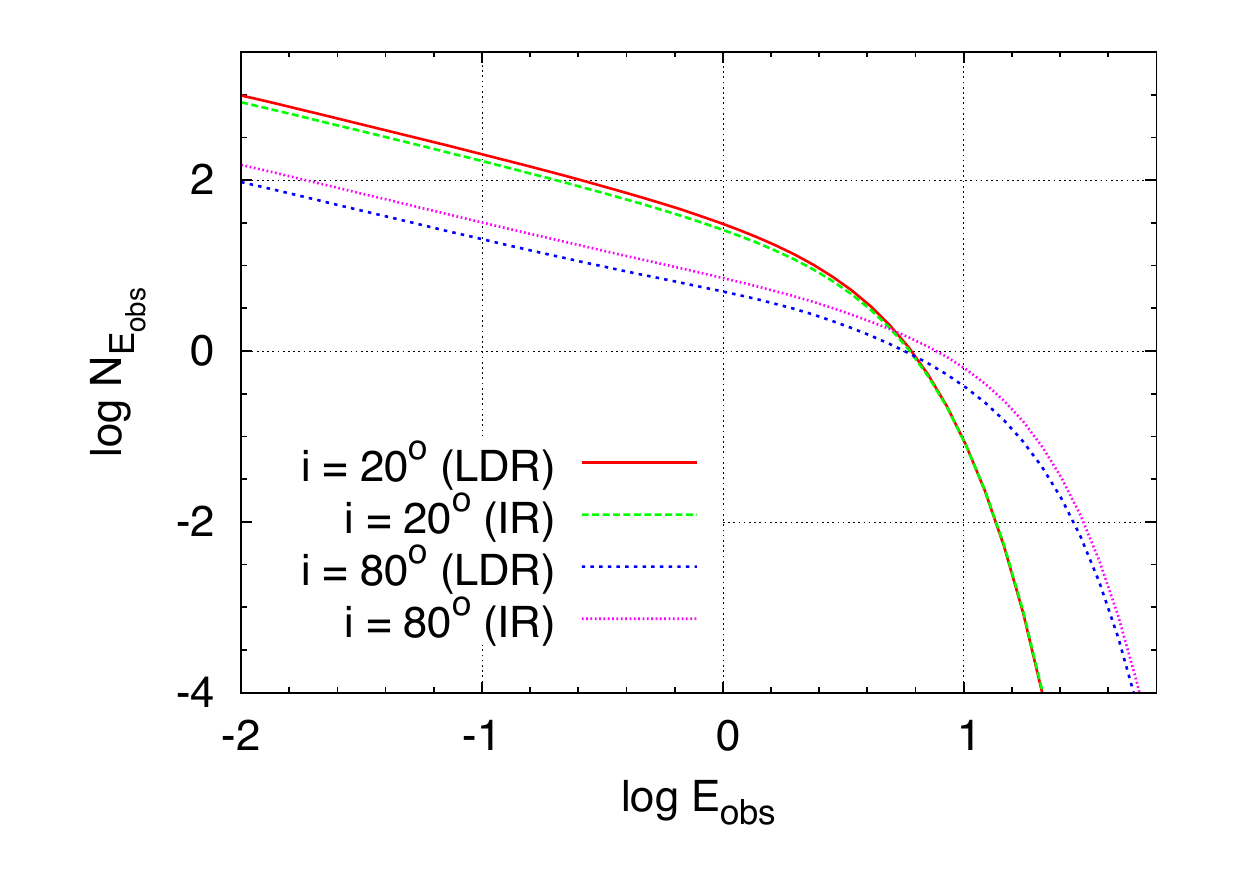}
\includegraphics[type=pdf,ext=.pdf,read=.pdf,width=8cm]{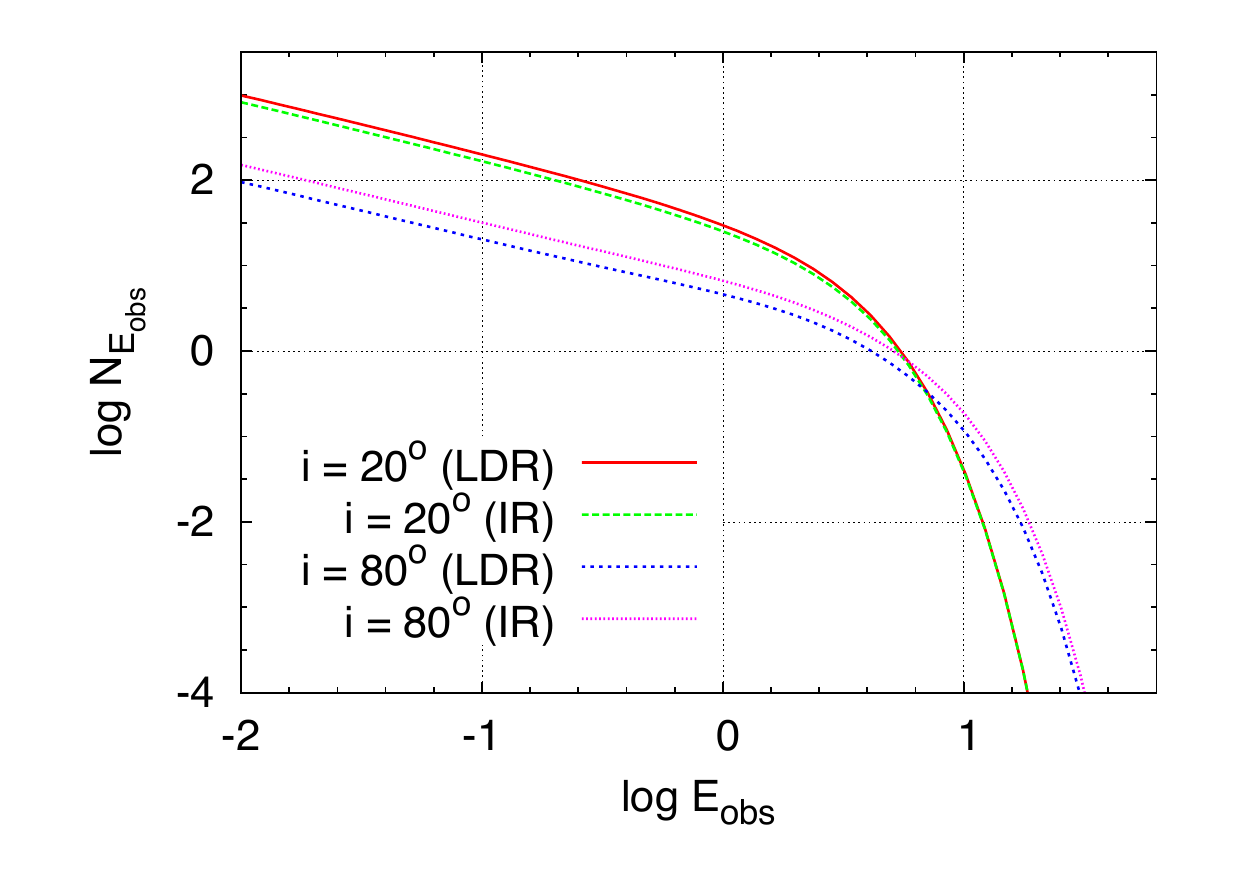}
\end{center}
\vspace{-0.6cm}
\caption{Effect of limb-darkened emission on the observed spectrum. Left panel: 
Kerr space-time with $a/M = 0.999$ and viewing angle $i = 20^\circ$ and $80^\circ$. 
Right panel: Johannsen-Psaltis space-time with $a/M = 1.1$, $\epsilon_3 = -0.5$, 
and viewing angle $i = 20^\circ$ and $80^\circ$. The other parameters are 
$M = 10$~$M_\odot$, $\dot{M} = 10^{19}$~g~s$^{-1}$, $D = 10$~kpc, and 
$f_{\rm col} = 1.5$. LDR/IR = limb-darkened/isotropic radiation. The outer radius 
of the accretion disk is at $r_{\rm out} = 10^5$~$M$. The left panel can be
compared with Fig.~9 of \citet{li05}. Flux density $N_{E_{\rm obs}}$ in 
$\gamma$ keV$^{-1}$ cm$^{-2}$ s$^{-1}$; photon energy $E_{\rm obs}$ in keV.}
\label{f-limb}
\end{figure}

\begin{figure}
\par
\begin{center}  
\includegraphics[type=pdf,ext=.pdf,read=.pdf,width=8cm]{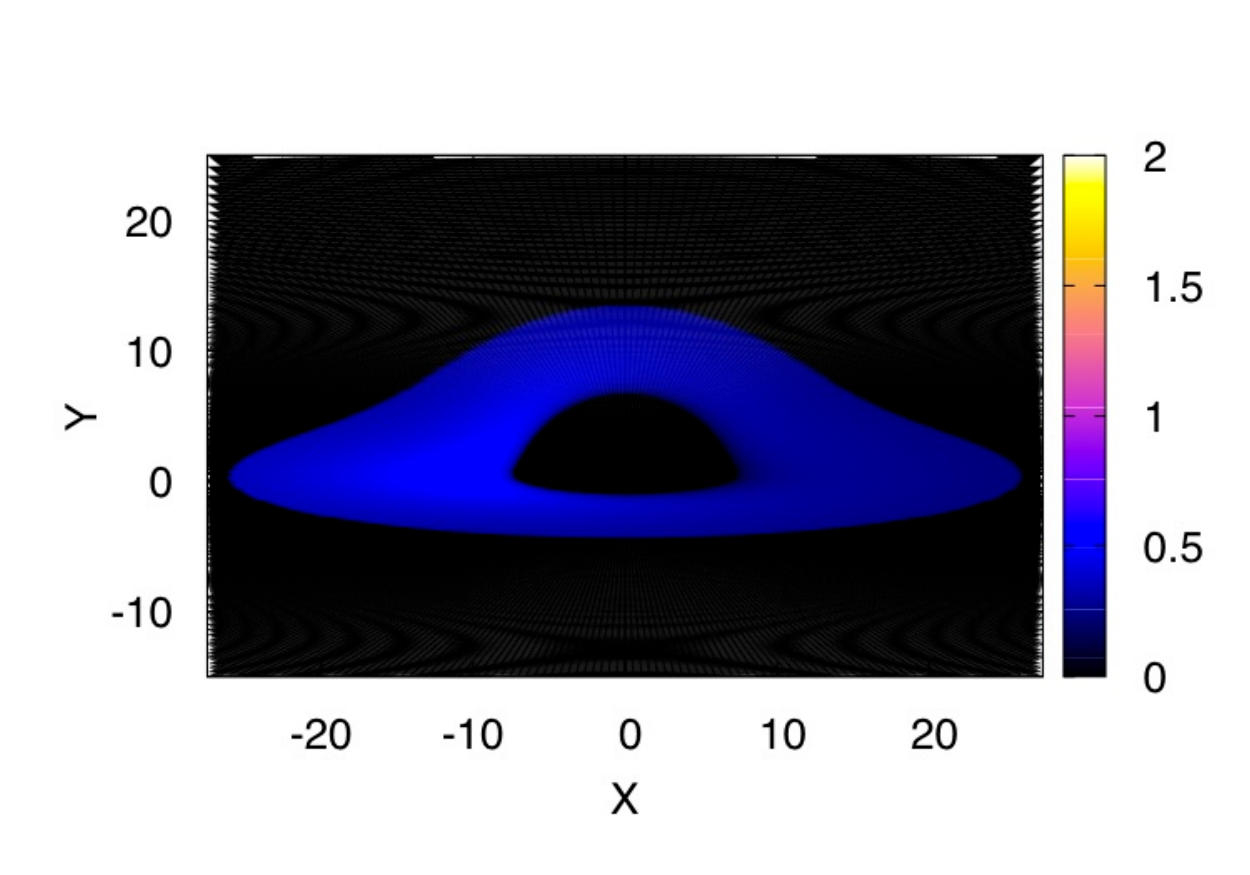}
\includegraphics[type=pdf,ext=.pdf,read=.pdf,width=8cm]{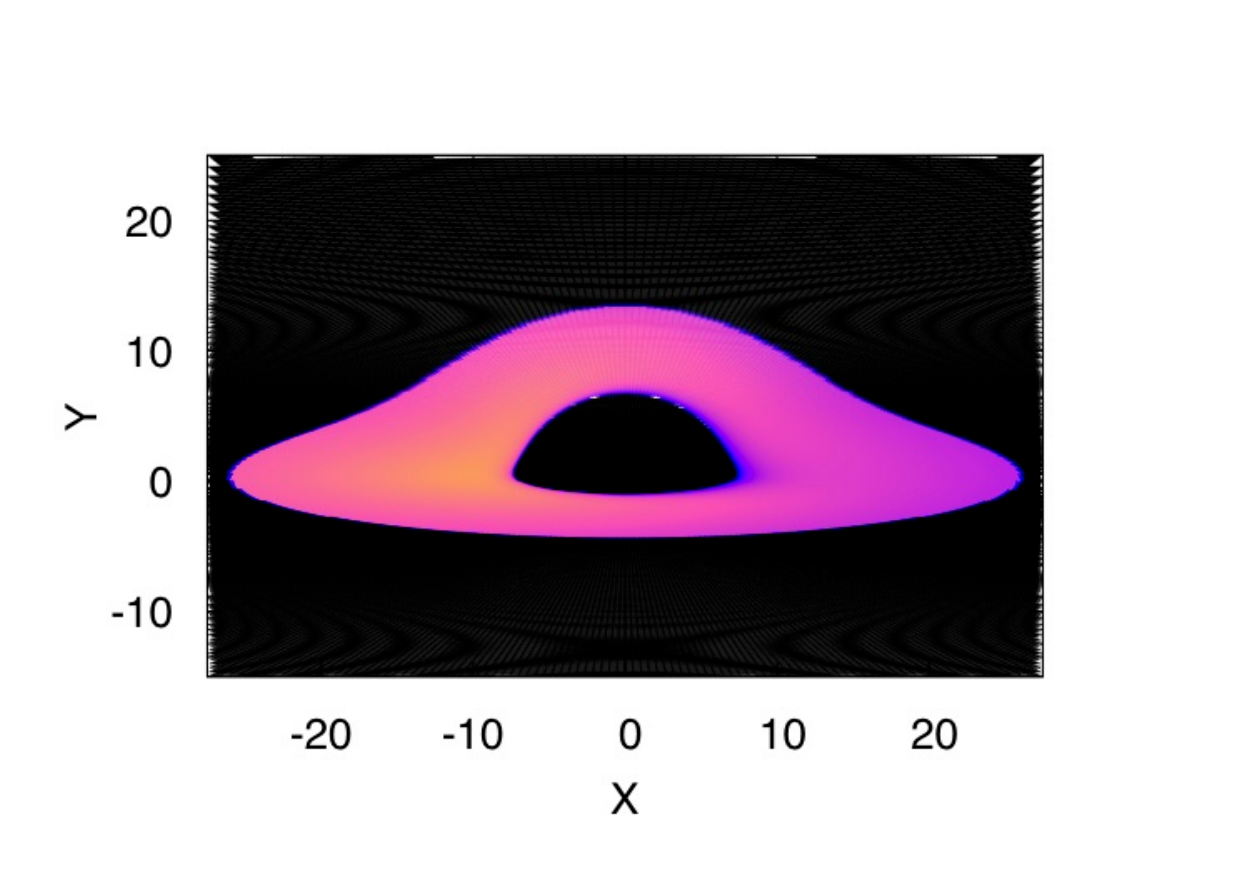} \\ \vspace{-0.3cm}
\includegraphics[type=pdf,ext=.pdf,read=.pdf,width=8cm]{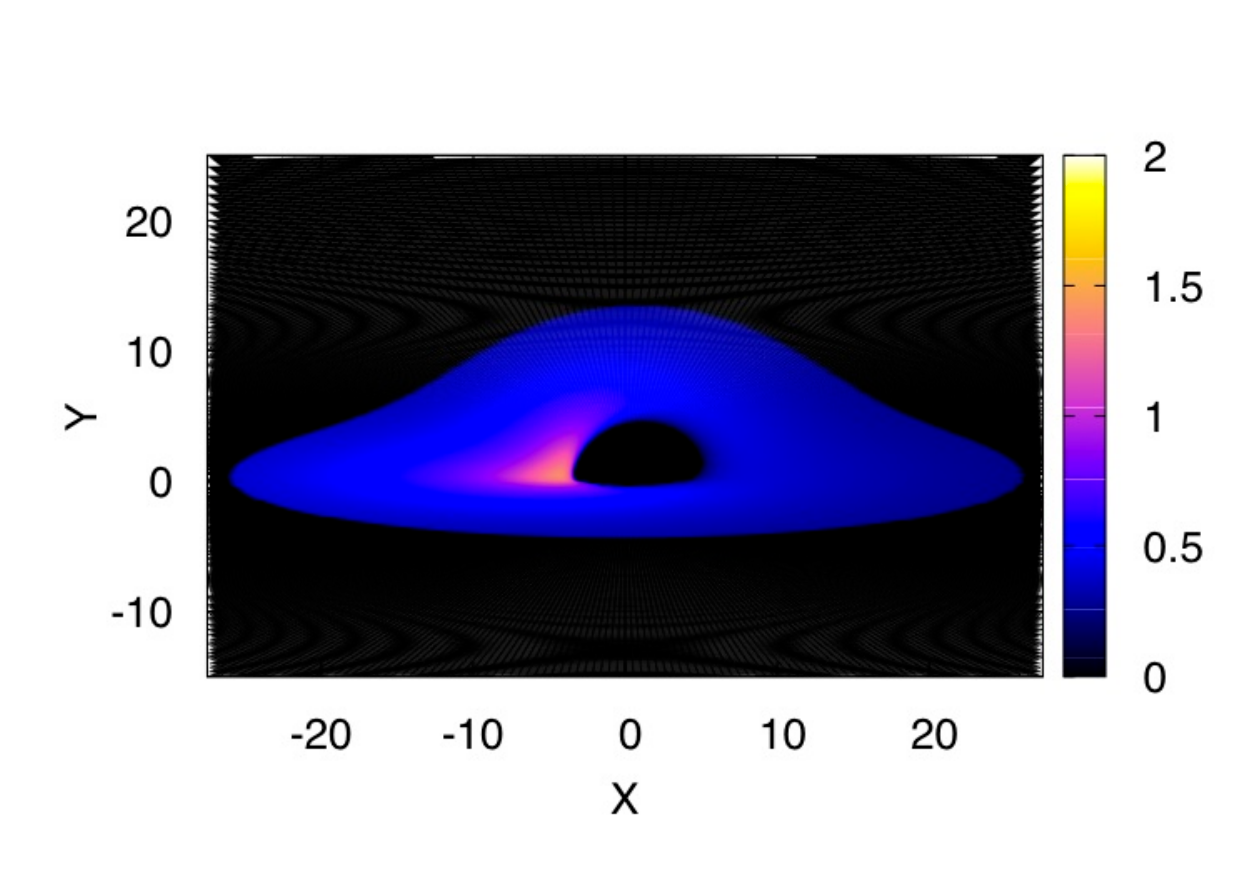}
\includegraphics[type=pdf,ext=.pdf,read=.pdf,width=8cm]{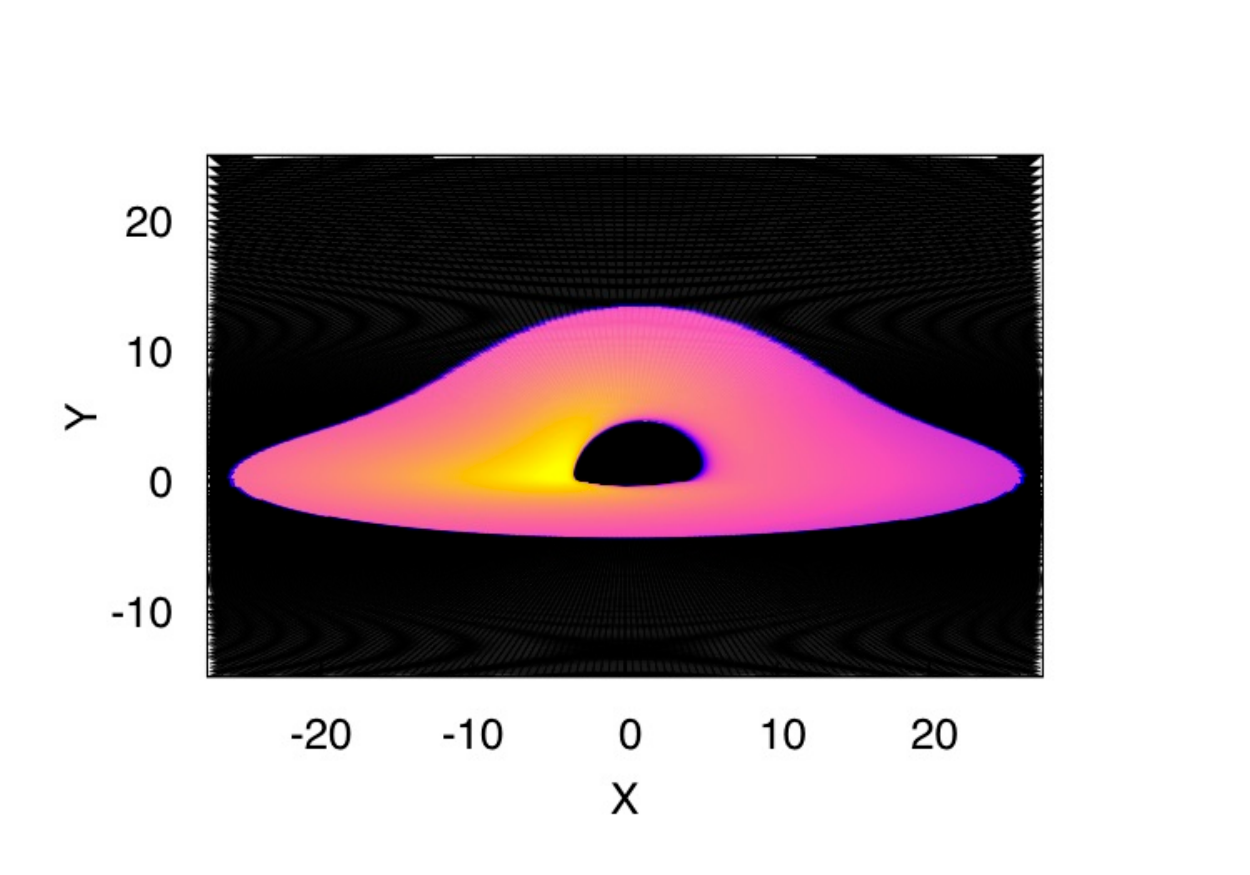}
\end{center}
\par
\vspace{-0.6cm}
\caption{Direct image of the accretion disk. Observed blackbody temperature
$T_{\rm obs}$ (left panels) and observed flux $\mathcal{F}_{\rm obs}$ (right 
panels) in Kerr space-time with spin parameter $a/M = 0$ (top panels) and 0.9 
(bottom panels). The other parameters are $M = 10$~$M_\odot$, 
$\dot{M} = 10^{18}$~g~s$^{-1}$, $i = 80^\circ$, and $f_{\rm col} = 1.6$. 
The outer radius of the accretion disk is $r_{\rm out} = 25$~$M$. 
$T_{\rm obs}$ in keV; $\mathcal{F}_{\rm obs}$ in arbitrary units
and logarithmic scale.}
\label{f-im1}
\end{figure}

\begin{figure}
\begin{center}  
\includegraphics[type=pdf,ext=.pdf,read=.pdf,width=8cm]{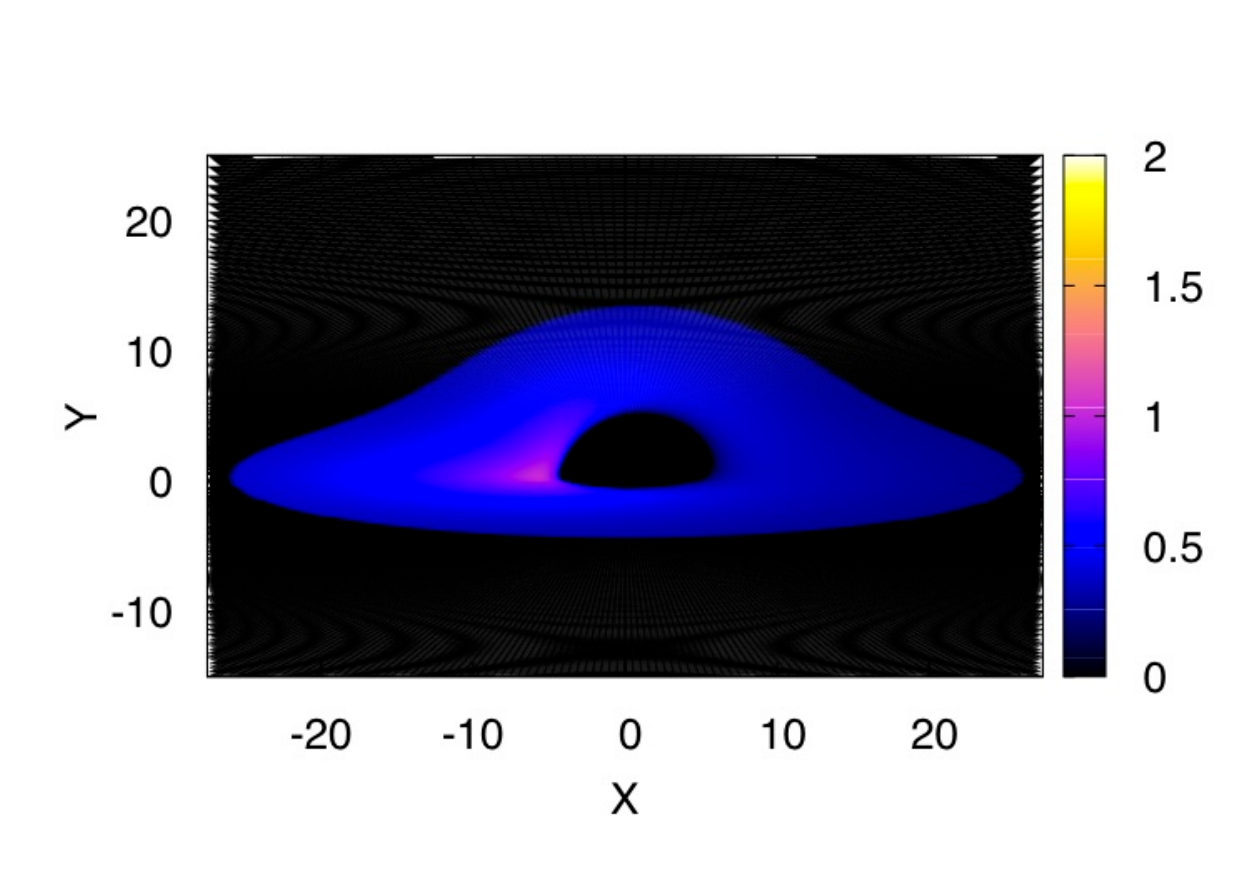}
\includegraphics[type=pdf,ext=.pdf,read=.pdf,width=8cm]{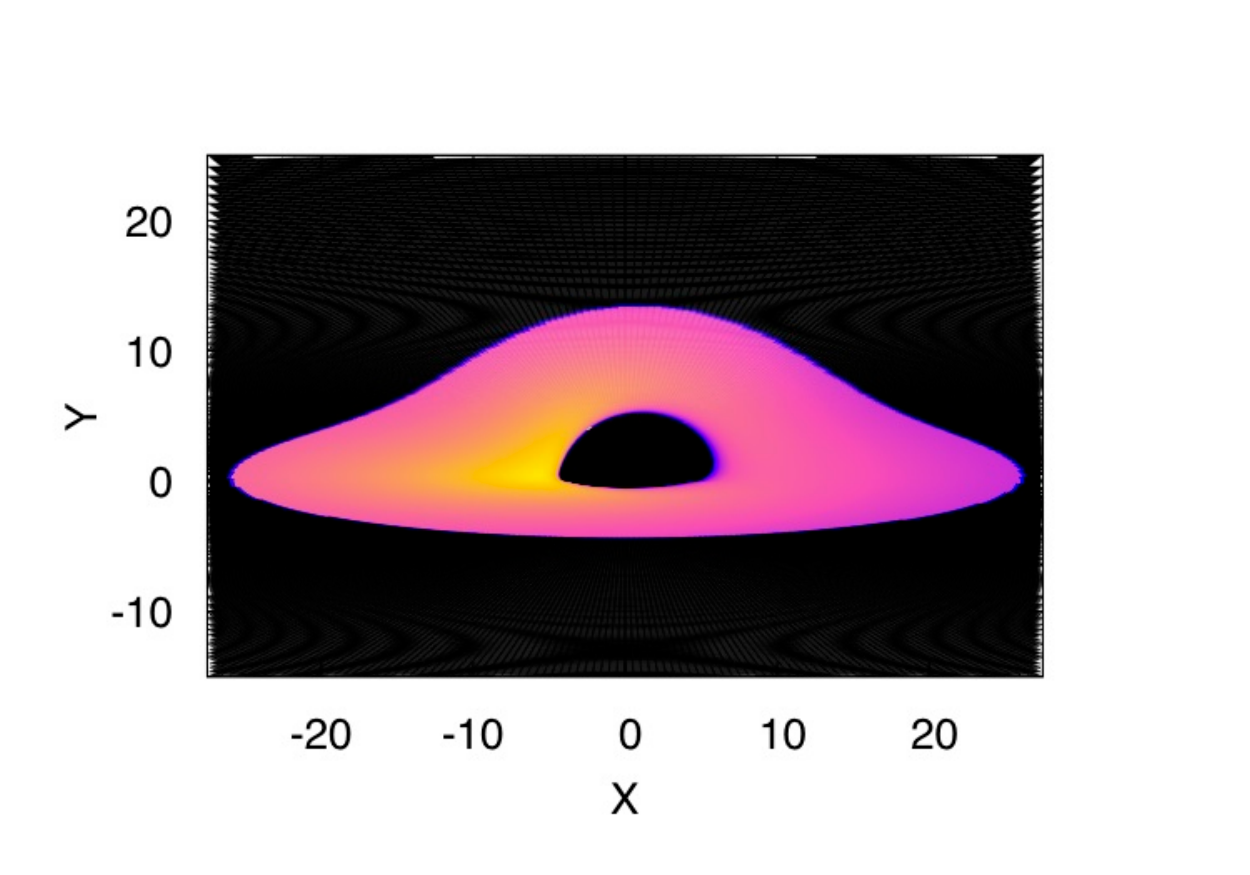} \\ \vspace{-0.3cm}
\includegraphics[type=pdf,ext=.pdf,read=.pdf,width=8cm]{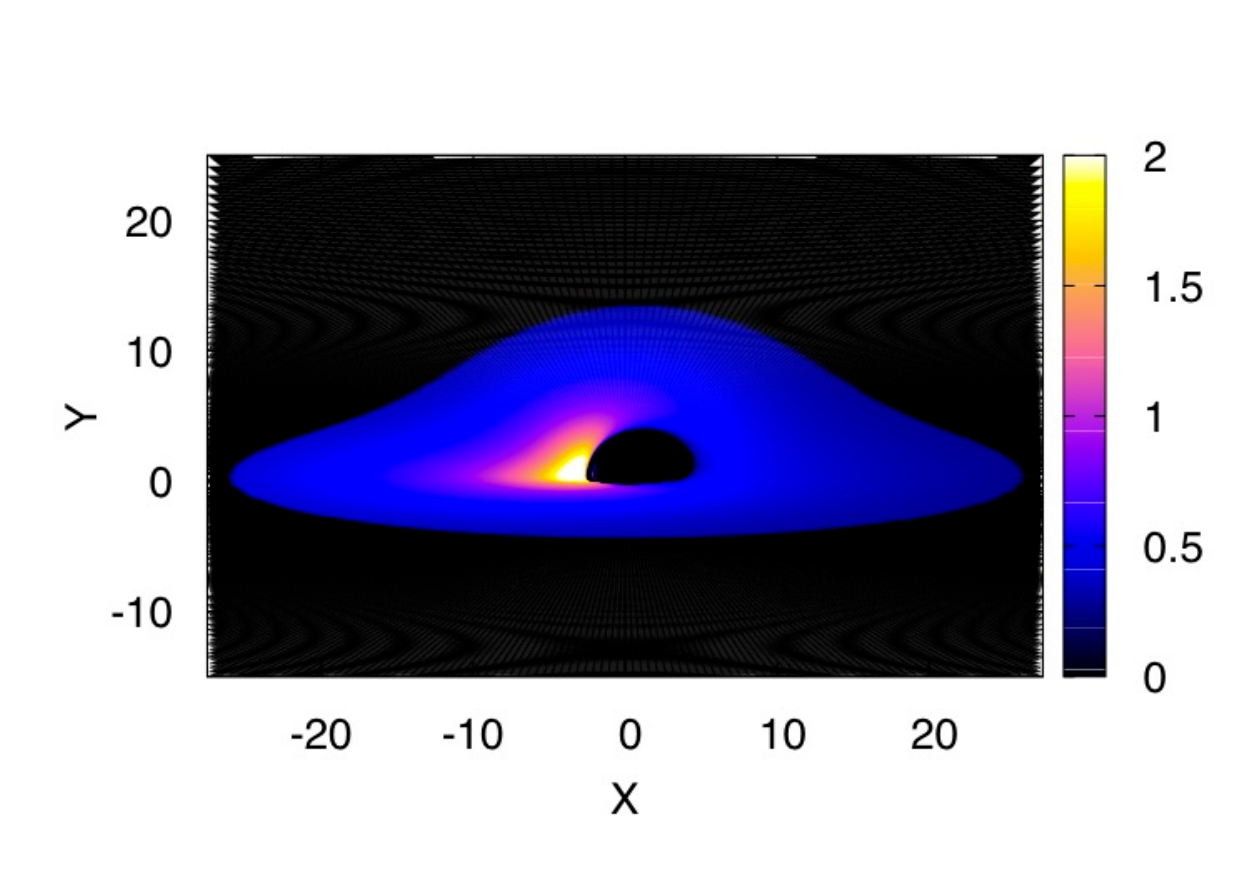}
\includegraphics[type=pdf,ext=.pdf,read=.pdf,width=8cm]{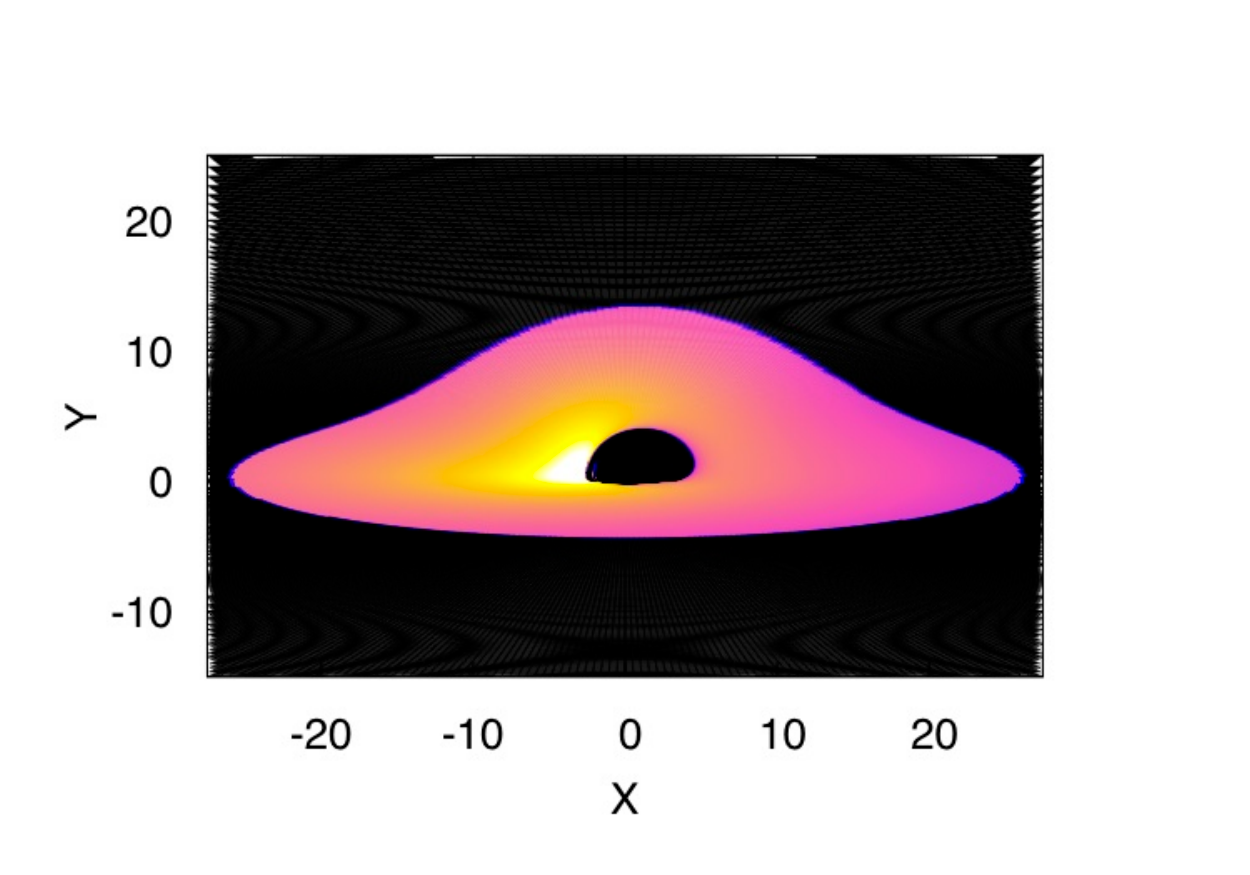}
\end{center}
\vspace{-0.6cm}
\caption{Direct image of the accretion disk. Observed blackbody temperature
$T_{\rm obs}$ (left panels) and observed flux $\mathcal{F}_{\rm obs}$ (right 
panels) in Johannsen-Psaltis space-time with spin parameter $a/M = 0.9$ and
deformation parameter $\epsilon_3 = -1$ (top panels) and 1 (bottom panels). 
The other parameters are $M = 10$~$M_\odot$, $\dot{M} = 10^{18}$~g~s$^{-1}$, 
$i = 80^\circ$, and $f_{\rm col} = 1.6$. The outer radius of the accretion 
disk is $r_{\rm out} = 25$~$M$. $T_{\rm obs}$ in keV; $\mathcal{F}_{\rm obs}$ 
in arbitrary units and logarithmic scale.}
\label{f-im2}
\end{figure}

\begin{figure}
\begin{center}  
\includegraphics[type=pdf,ext=.pdf,read=.pdf,width=8cm]{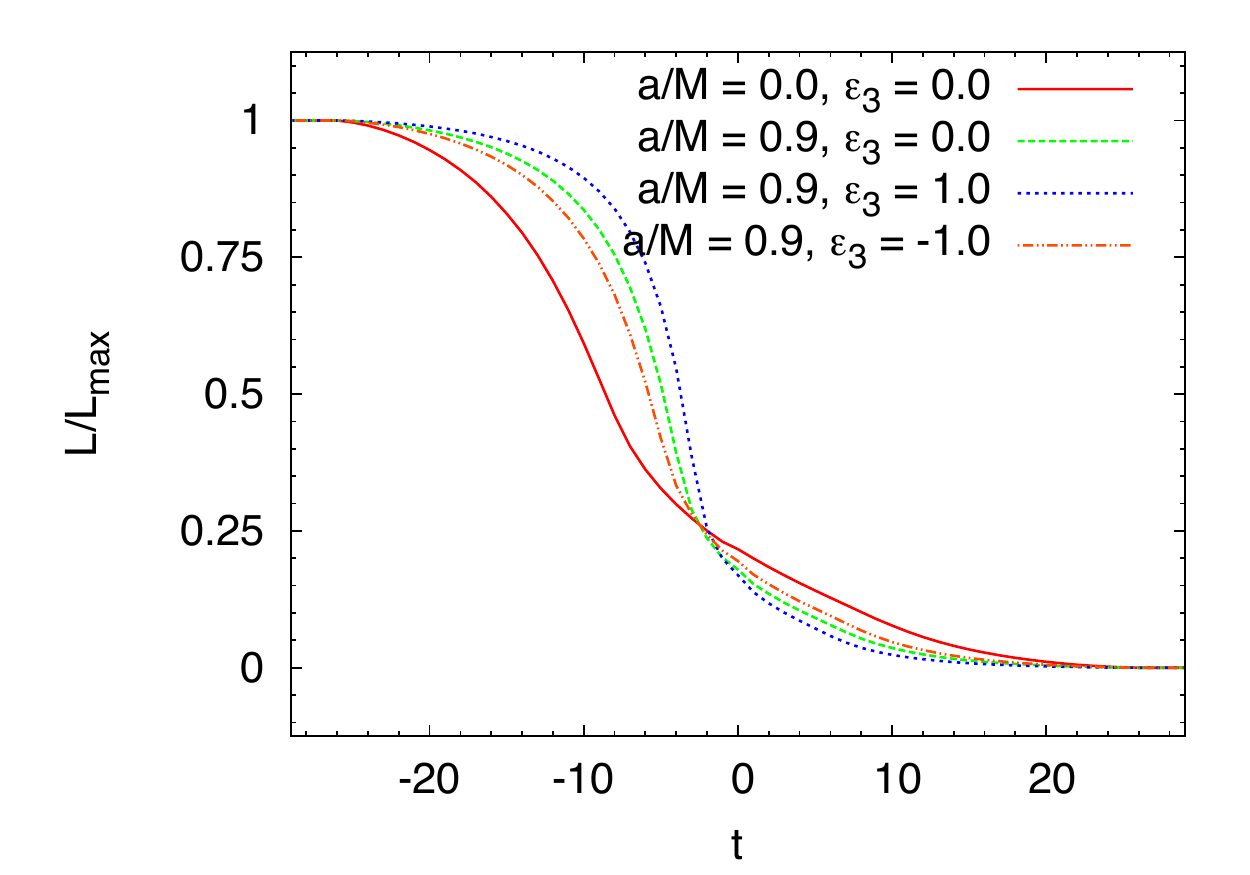}
\includegraphics[type=pdf,ext=.pdf,read=.pdf,width=8cm]{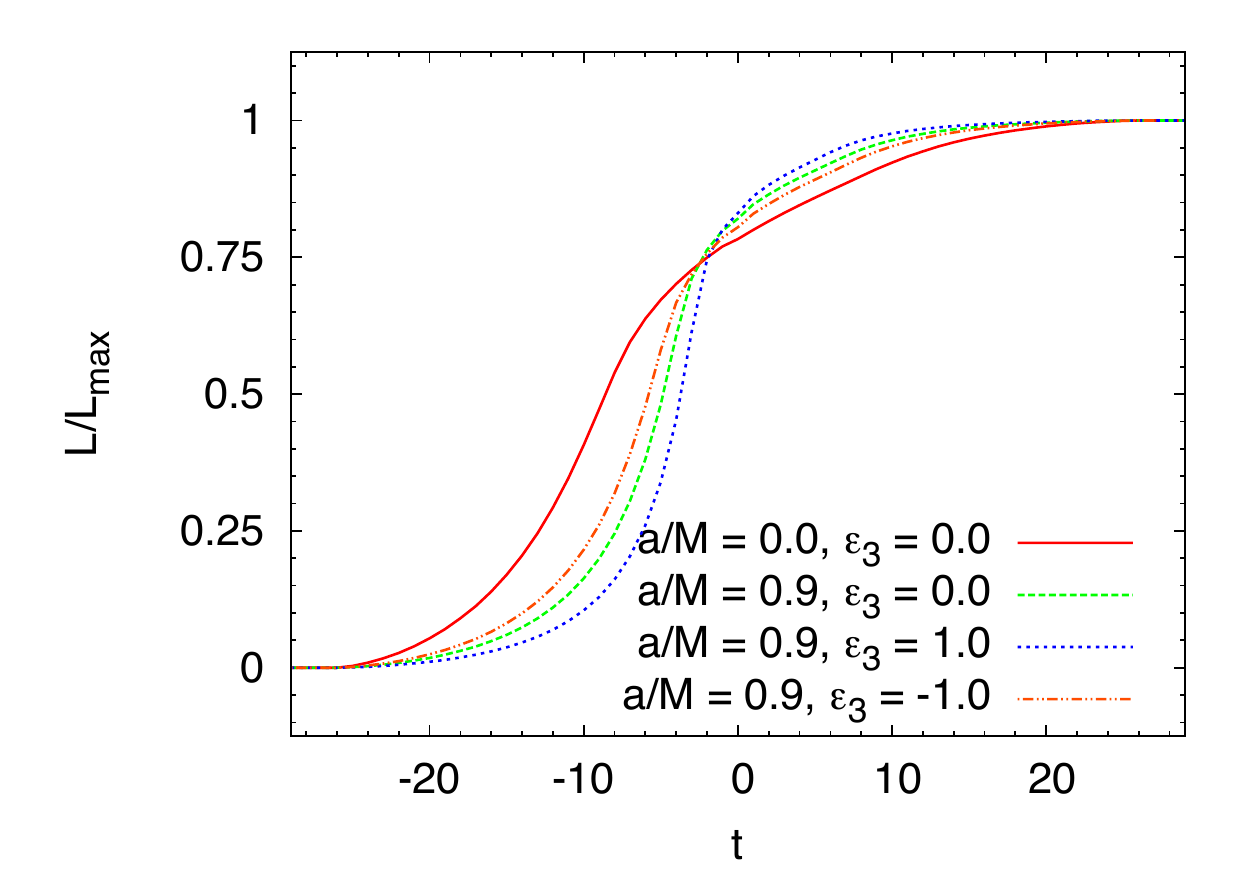}
\end{center}
\vspace{-0.6cm}
\caption{Light curves of ingress (left panel) and egress (right panel) 
during an eclipse for the cases of Figs.~\ref{f-im1} and \ref{f-im2}. Time $t$ in 
units of $M/v_{\rm c}$, where $M$ is the mass of the black hole candidate 
and $v_{\rm c}$ is the velocity of the stellar companion. See the text for details.}
\label{f-ecl}
\end{figure}

The properties of JP BHs with $M$, $a$, and $\epsilon_3$ have been quite discussed in 
the recent literature \citep{b11a3,b12agn,b12a,b12b,b12c,leo1,leo2,jpk,chen1,chen2,kraw}. 
One can note that these objects may have properties  fundamentally different from the
ones expected for Kerr BHs. For instance, when the gravitational
force on the equatorial plane turns out to be stronger/weaker than
the Kerr case (this corresponds to $\epsilon_3 <
0 / \epsilon_3 > 0$), the maximum value of $a/M$ can presumably be larger/smaller
than 1, and increases/decreases as $\epsilon_3$ decreases/increases \citep{b11a3}.
Another important feature to bear in mind is that the ISCO may not be determined
by the orbital stability along the radial direction, as in Kerr, but it may be marginally
vertically stable. The ISCO radius should thus be computed as described in
\citet{bb11,bb11isco}. The event horizon of JP BHs has been discussed in 
\citet{leo1} and the radius $r_{\rm H}$ can be found from\footnote{Let us note that there 
are a few definitions of horizon. Here we are talking about the {\it event horizon}, i.e. a 
boundary in the space-time beyond which events cannot affect an outside 
observer. In a stationary space-time, the event horizon is also an {\it apparent horizon}, 
which is a surface of zero expansion for a congruence of outgoing null geodesics 
orthogonal to the surface. This means that at the apparent horizon null geodesics 
must have $dr/dt = 0$, which implies $g^{rr} = 0$, see e.g. \citet{poisson}.
Eq.~(\ref{eq-horizon}) follows from $g^{rr} = 0$ in the case of the JP metric. 
The horizon relevant for the black hole thermodynamics is instead the {\it Killing horizon}, 
which is a null hyper-surface on which there is a null Killing vector field. For the 
metric in~(\ref{eq-jp}), the Killing horizon is defined by $g_{tt} g_{\phi\phi} - g_{t\phi}^2 = 0$.
When the Hawking's rigidity theorem can be applied (like in the Kerr space-time), the 
event horizon and the Killing horizon coincide \citep{hawking}. However, in 
general that is not true.} 
\be\label{eq-horizon}
\Delta + a^2 h \sin^2 \theta = 0 \, .
\ee
For non-vanishing $a$ and $h$, the radius $r_{\rm H}$ is a function of the angle $\theta$.
It is also possible to show that the topology of the event horizon of these BHs may be
non-trivial and that BHs with a topologically non-trivial event horizon may be created
by astrophysical processes like the accretion from disk \citep{leo1}. The fundamental
properties of the JP space-times shown in Figs.~\ref{f-iobs}-\ref{f-ecl} are reported in 
Tab.~\ref{tab2}.

Examples of the observed spectrum of thin accretion disks in the
JP background are shown in Fig.~\ref{f-iobs}, for $a/M = 0.8$
and $\epsilon_3 = 2$ (left panel) and for $a/M = 1.2$ and
$\epsilon_3 = -1$ (right panel). 
Fig.~\ref{f-iobs} shows the effect of light bending in the JP background.
As in the Kerr case, the effect is more and more important as the
disk's inclination angle increases and the observer approaches
the equatorial plane of the system.
Fig.~\ref{f-limb} shows instead the effect on the observed spectrum
of the properties of the emission of the accretion disk. It compares the
cases of isotropic ($\Upsilon = 1$) and limb-darkened emission
($\Upsilon = \frac{1}{2} + \frac{3}{4} \cos\xi$, $\xi$ being the angle 
between the wavevector of the photon emitted by the disk and the 
normal of the disk surface). The left panel of Fig.~\ref{f-limb} can be
compared with Fig.~9 of \citet{li05}. As noticed in~\citet{li05},
for edge-on disks the effect of limb-darkening cannot be absorbed
by a redefinition of $\dot{M}$ and $f_{\rm col}$; the spectrum has 
indeed a more pronounced hump before the exponential cut-off.
The case of non-Kerr background does not introduce any new
qualitatively different feature with respect to the Kerr case already 
discussed in the literature.

\section{Other observable features}

In the previous sections, I showed how the code can compute the thermal
spectrum of a geometrically thin and optically thick accretion disk in a
generic space-time, comparing the results for the Kerr case with the ones
presented in \citet{li05}. While even the continuum-fitting method
is not immune to criticism, this approach to test the nature of astrophysical
BH candidates is likely the best we can do with current observational data 
and theoretical knowledge \citep{cfm}. Constraints on the spin parameter-deformation
parameter plane obtained with the code by analyzing the data of specific 
sources in the high-soft state will be presented in a companion paper \citep{bgs}.
In this section, I am going to show some simple extensions of the code:
the computation of the direct image of the disk and the prediction of the
light curve during an eclipse. Current facilities cannot observe these features,
but future X-ray experiments may do the job and provide additional information
about the geometry of the space-time around BH candidates.

\subsection{Direct image of thin accretion disks}

In the case of a geometrically thick and optically thin accretion disk
around a BH candidate, it should be possible to observe the so-called
BH shadow; that is, a dark area over a brighter background. As the
exact shape of the shadow is determined exclusively by the geometry
of the space-time, its detection can be used to investigate the nature
of the BH candidate \citep{ktf,naoki,jp10,leo2}. In the case of geometrically
thin and optically thick accretion disks, the image turns out to be quite
different. For Schwarzschild and Kerr BHs, it has been already discussed
in the literature \citep{luminet,fukue1,fukue2,rohta}. The generalization
to non-Kerr space-times is straightforward (see also \citet{kraw}). The 
temperature and the photon flux as detected by a distant observer are given by
\be\label{eq-im-1}
T_{\rm obs} &=& g T_{\rm col} \, , \\
\mathcal{F}_{\rm obs} &=& g^4 \mathcal{F} \, , \label{eq-im-2}
\ee
where $g$ is the redshift factor given by Eq.~(\ref{eq-red}). Eq.~(\ref{eq-im-1})
follows from the definition of $g$ and from Eq.~(\ref{eq-i-bb}), while Eq.~(\ref{eq-im-2})
is a consequence of the Liouville's theorem $I_{\rm e}(\nu_{\rm e})/\nu_{\rm e}^3 
= I_{\rm obs} (\nu_{\rm obs})/\nu^3$. Fig.~\ref{f-im1} shows the direct
image of $T_{\rm obs}$ and $\mathcal{F}_{\rm obs}$ for a Schwarzcshild
BH (top panels) and for a Kerr BH with $a/M = 0.9$ (bottom panels), in the
case of an observer with inclination angle of $80^\circ$. The asymmetry
with respect to the $X$ axes is due to the effect of light bending, while
the one with respect to the $Y$ axes comes from the Doppler boosting.
Fig.~\ref{f-im2} shows the same images in the case the accretion disk
is in the JP space-time with spin $a/M = 0.9$ and deformation parameter $\epsilon_3 = -1$
(top panels) and 1 (bottom panels). As we can see, the shape of the central
dark area and the intensity map of the image depend clearly on the 
geometry of the space-time around the compact object. Such images are 
definitively impossible to observe with current X-ray facilities, but they may 
be seen with future X-ray interferometry techniques.

\subsection{Light curve during an eclipse}

The observation of the light curve when the accretion disk around a BH 
candidate is eclipsed by the stellar companion can also provide information
about the geometry of the space-time around the compact object \citep{fukue1}.
While current observational facilities would already have the correct time 
resolution to observe this feature, so far we know only one binary system
with a BH candidate showing an eclipse in the X-ray spectrum. This source
is M33~X-7, which is unfortunately quite far and dim, and present detectors
cannot measure its light curve with an accuracy to observe any relativistic 
effect \citep{m33ec}.

Fig.~\ref{f-ecl} shows the light curves of the ingress (left panel) and egress 
(right panel) during an eclipse for the four cases shown in Figs.~\ref{f-im1} 
and \ref{f-im2}. The inclination angle of the observer is still $i = 80^\circ$. 
In these simulations, I assume that the angular momentum of the companion 
star is parallel to the one of the accretion disk, as it should be more likely 
expected from considerations on the evolution of the system. That means 
that the companion star occults first the approaching (blue-shifted) part of 
the disk and then the receding (red-shifted) one. The companion star is 
simply modeled as an obstacle with vertical edge and its atmosphere is 
completely neglected, so the disk's luminosity at the time $t$ is
\be
L(t) = \int h_i (X - v_{\rm c} t) I_{\rm obs}(\nu) \frac{dX dY}{D^2} \, ,
\ee
where $i = {\rm ingress}$ or regress, $v_{\rm c}$ is the velocity of the stellar 
companion, and
\be
h_{\rm ingress} (X - v_{\rm c} t) = \left\{
\begin{array}{rl}
0 & \text{if } X - v_{\rm c} t \le 0 \, , \\
1 & \text{if } X - v_{\rm c} t > 0 \, .
\end{array} \right. \\
h_{\rm egress} (X - v_{\rm c} t) = \left\{
\begin{array}{rl}
1 & \text{if } X - v_{\rm c} t < 0 \, , \\
0 & \text{if } X - v_{\rm c} t \ge 0 \, .
\end{array} \right.
\ee
As the shape of the light curve depends on the 
background metric, the possible detection of this feature can be used to 
constrain possible deviations from the Kerr nature of a BH candidate.
However, in order to extract information on the space-time geometry from 
real data, a reliable atmospheric model would be necessary \citep{rohta2}.

If we could get accurate images of the disk, like the ones shown in Figs~\ref{f-im1} 
and \ref{f-im2}, we could determine at the same time $a$ and $\epsilon_3$.
However, that is surely out of reach in the near future. In the case of the light
curve during an X-ray eclipse, basically we measure the slope of the curve
and the asymmetry between ingress and egress. Assuming a Kerr background,
we could immediately estimate $a$ (assuming $M$ is known independently).
If we are going to test the Kerr-nature of the BH candidate, we find a degeneracy
between $a$ and $\epsilon_3$. However, the light curve is essentially sensitive only to
the effect of Doppler boosting; that is, the asymmetry of the apparent image of the
disk with respect to the $X = 0$ axes. It is not very sensitive to the effect of 
light bending (responsible to the asymmetry with respect to the $Y = 0$ axes)
and to the gravitational redshift (which produces corrections symmetric with respect
to the $X = 0$ axes). The combination of the fit of the thermal spectrum and of
the light curve during an eclipse for the same BH candidate should thus break
the degeneracy between $a$ and $\epsilon_3$ and allows for the identification
of a limited allowed region in the spin parameter-deformation parameter plane.

\section{Conclusions}

General Relativity has been tested and verified in Earth's gravitational
field, in the Solar System, and by studying the motion of binary 
pulsars. Thanks to recent theoretical progresses and high-quality data from
present and near-future observational facilities, it is now possible to start 
testing the theory in the strong field regime. One of the most intriguing 
predictions of General Relativity is that the final product of the gravitational 
collapse is a Kerr black hole: a very simple object completely characterized 
by only two quantities (the mass $M$ and the spin angular momentum $J$) 
in a very specific way. The study of the thermal spectrum of geometrically
thin accretion disks around stellar-mass black hole candidates can provide 
information about the geometry of the space-time and can thus be used to
check the Kerr black hole paradigm.

In this paper, I have presented a code based on a ray-tracing approach
and designed to compute the thermal emission of thin accretion disks
around generic compact objects. In particular, the code can compute the
thermal spectrum of a thin disk around a compact object with mass $M$,
spin parameter $a$, and a deformation parameter which measures 
possible deviations from the Kerr background. By comparing these
theoretical predictions with X-ray data of stellar-mass BH candidates
in the high-soft state, one can constrain the nature of the compact
object in the spin parameter-deformation parameter plane. Constraints
from specific sources will be presented in a forthcoming paper \citep{bgs}.
The code can also be used to compute other observational features
of a thin disk, like its direct image (Figs.~\ref{f-im1} and \ref{f-im2}) or its 
light curve during an eclipse (Fig.~\ref{f-ecl}), which may be observed
with future X-ray facilities.


\begin{acknowledgments}
I would like to thank Luca Maccione for fixing a bug in the code.
This work was supported by the Humboldt Foundation, Fudan 
University, and the Thousand Young Talents Program.
\end{acknowledgments}


\end{document}